\documentclass[%
 reprint,
 amsmath,
 amssymb,
 aps
]{revtex4-1}

\usepackage{graphicx}
\usepackage{dcolumn}
\usepackage{bm}

\usepackage{mathtools}
\usepackage{float}
\usepackage{siunitx}

\begin{document}

\title{BEAM BREAK-UP CURRENT LIMIT IN MULTI-TURN ERLs AND CBETA}

\thanks{This work was performed with the support of NYSERDA (New York State Energy Research and Development Agency).}
\author{W.~Lou}
\email{wl528@cornell.edu}
\author{G.H.~Hoffstaetter}
\affiliation{CLASSE, Cornell University, Ithaca, NY 14853, USA
}

\date{\today}


\begin{abstract}
This paper uses theory and simulation of the Beam Break-Up instability (BBU) for multi-turn ERLs to determine and to optimize the current limit of the Cornell Brookhaven Energy-Recovery-Linac Test Accelerator (CBETA). Currently under construction at Cornell University's Wilson Laboratory, the primary structures of CBETA for beam recirculation include the Main Linac Cryomodule and the Fixed Field Alternating Gradient beamline. As the electron bunches pass through the MLC cavities, Higher Order Modes (HOMs) are excited. The recirculating bunches return to the cavities to further excite HOMs, and this feedback loop can give rise to BBU. We will first explain how BBU effect is simulated using the tracking software BMAD, and check the agreement with the BBU theory for the most instructive cases. We then present simulation results on how BBU limits the maximum achievable current of CBETA with different HOM spectra in the cavities. Lastly we investigate ways to improve the threshold current of CBETA. 
\end{abstract}

\maketitle

\section{Introduction}

Energy recovery linacs (ERLs) open up a new regime of beam parameters with large current and simultaneously small emittances, bunch lengths, and energy spread. The Cornell BNL ERL Test Accelerator (CBETA) is the first accelerator that is constructed to analyze the potential of multi-pass ERLs with superconducting SRF accelerating cavities \cite{cbetacdr}. New beam parameters of ERLs allow for new experiments such as nuclear and high energy colliders, electron coolers, internal scattering experiments, X-ray sources or Compton backscattering sources for nuclear or X-ray physics \cite{exp1}\cite{exp2}\cite{exp3}.
By recirculating charged beams back into the accelerating cavities, energy can be recovered from the beams to the electromagnetic fields of the cavities. Energy recovery allows an ERL to operate at a much higher current than conventional linacs, where the current is limited by the power consumption by the cavities. While electron beams recirculate for thousands of turns in storage rings, they travel only a few turns in an ERL before being dumped. The short circulation time allows beam emittances to be as small as for a linac. The potentials for high beam current with simultaneously low emittances allows an ERL to deliver unprecedented beam parameters.

CBETA is currently under construction at Cornell University's Wilson Laboratory. This is a collaboration with BNL, and will be the first multipass ERL with a Fixed Field Alternating (FFA) lattice. It serves as a prototype accelerator for electron coolers of Electron Ion Colliders (EICs). Both EIC projects in the US, eRHIC at BNL and JLEIC at TJNAF will benefit from this new accelerator \cite{ipac2017}.

\begin{figure}[!htb]
   \centering
   \includegraphics*[width=235pt]{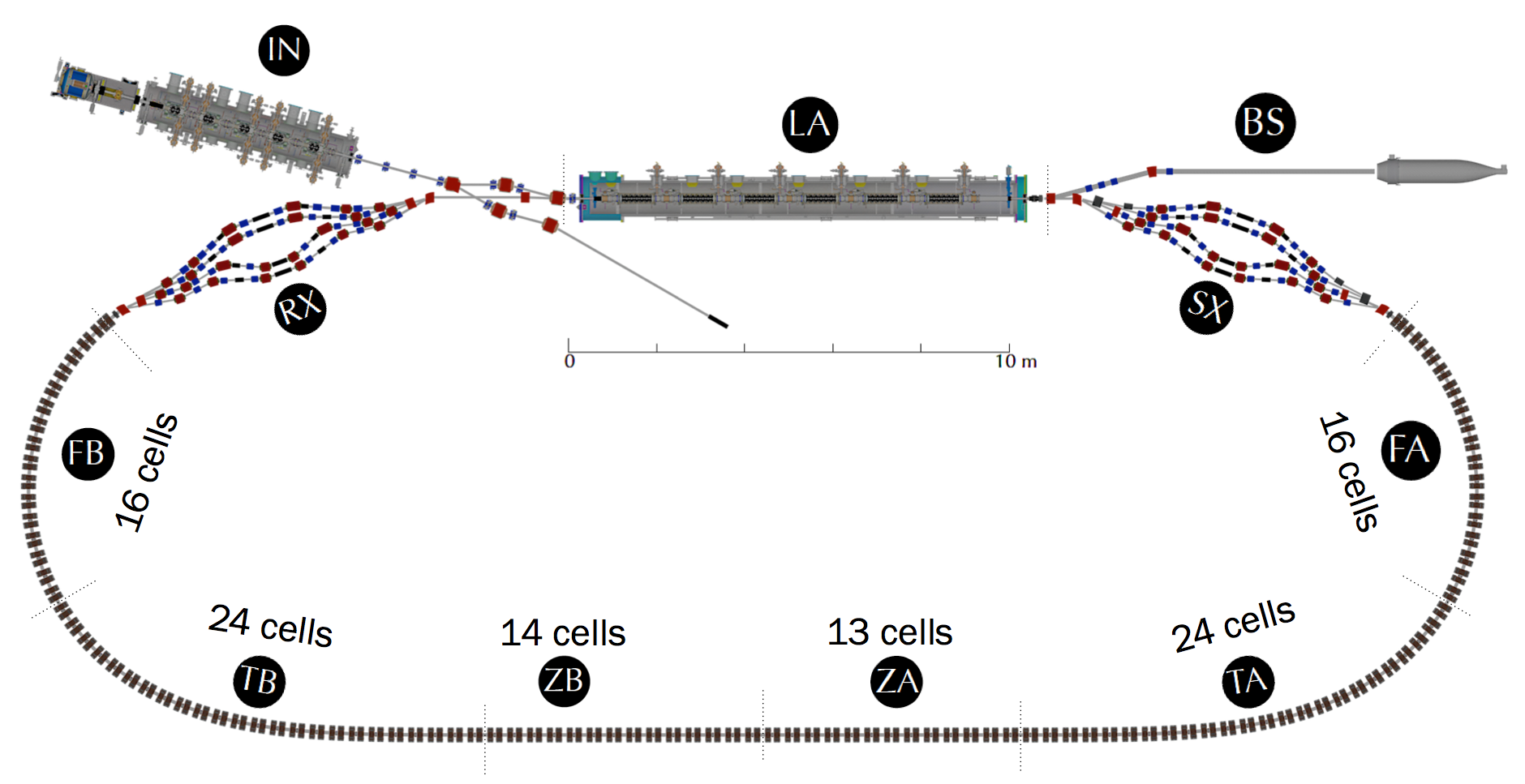}
   \caption{Layout of CBETA. The sections labeled (IN) and (LA) are the injector and MLC cavities respectively. Sections (FA), (TA), (ZA), (ZB), (TB), and (FB) form the FFA beamline which can accommodate four recirculating orbits with energy ranging from 42 MeV to 150 MeV. Sections (SX) and (RX) are splitters and recombiners which control the path-length of each recirculation pass.}
   \label{CBETA}
\end{figure}

Fig.~\ref{CBETA} shows the design layout of CBETA. At full operation, CBETA will be 4-pass ERL with maximum electron beam energy of 150 MeV. This is achieved by first accelerating the electron beam to 6~MeV by the injector (IN). The beam is then accelerated by the Main Linac Cryomodule (MLC) cavities (LA) four times to reach 150~MeV, then the beam is decelerated four times down to 6~MeV before stopped (BS). The beam passes through the MLC cavities for a total of eight times, each time with an energy gain of $\pm$36 MeV. The field energy in the cavities is transfered to the beam during acceleration, and recovered during deceleration. Transition from acceleration to deceleration is achieved by adjusting the path-length of the forth recirculation turn to be an odd multiple of half of the RF wavelength. The path-length of all the other turns is exactly an integer multiple of the RF wavelength. CBETA can also operate as a 3-pass, 2-pass, or 1-pass ERL with properly adjusted configuration.

While the beam current in ERLs is no longer limited by the power consumption in the cavities, there will be new, higher limits to the current. These are Higher Order Modes (HOMs) heating and the recirculative Beam Breakup (BBU) instability. 

BBU occurs in recirculating accelerators as the recirculated beam bunches interact with the HOMs in the accelerating cavities. The most relevant HOMs for BBU are the dipole HOMs which give a transverse kick to the bunches. The off-orbit bunches return to the same cavity and excite the dipole HOMs which can kick the subsequent bunches further in the same direction. The effect can build up and can eventually result in beam loss. With a larger beam current the effect becomes stronger, so BBU is a limiting factor on the maximum achievable current, called the threshold current $I_\text{th}$. With multiple recirculation passes, bunches interact with cavities for multiple times, and the $I_\text{th}$ can significantly decrease \cite{bbu_Georg_Ivan}. The low and high target currents of CBETA are 1~mA and 40~mA respectively, for both the 1-pass mode and 4-pass mode. Simulations are required to check whether the $I_\text{th}$ is above these target values.  

\section{BBU Simulation Overview}

Cornell University has developed a simulation software called BMAD to model relativistic beam dynamics in customized accelerator lattices \cite{BMAD}. Subroutines have been established to simulate specifically BBU and to find the $I_\text{th}$ for a specific lattice design. The program requires the lattice to have at least one recirculated cavity with at least one HOM assigned to it. There are six MLC cavities in the CBETA lattice, and multiple HOMs can be assigned to each cavity. The following two subsections describe how the HOM data are generated, and how BMAD finds the $I_\text{th}$.

\subsection{HOM simulation and assignment}
To run BBU simulation we must first obtain the HOM characteristics. Each HOM is characterized by its frequency~$f$, shunt impedance $(R/Q)$, quality factor $Q$, order $m$, and polarization angle $\theta$.
Since the MLC cavities have been built and commissioned, one would expect direct measurement of HOM spectra from the cavities. Unfortunately, the measured spectra contain hundreds of HOMs, and it is difficult to isolate each individual HOM and compute their characteristics, particularly $R/Q$. Therefore, instead of direct measurement, we simulate the HOM profiles using the known and modelled cavity structures \cite{Valles}. The simulation has been done using the CLANS2 program \cite{CLANS2}, which can model the fields and HOM spectrum within a cavity.

In reality each cavity is manufactured with small unknown errors. The cavity shape are characterized by ellipse parameters. The fabrication tolerance for the CBETA MLC cavities require the errors in these parameters to be within $\pm$\SI{125}{\micro\meter}. For simplicity we use $\epsilon$ to denote the maximum deviation ,i.e. $\epsilon$ = \SI{125}{\micro\meter} for realistic CBETA cavities. In the CLANS2 program, random errors are introduced to the modelled cavity shape within a specified $\epsilon$. The cavity is then compressed to obtain the desired fundamental accelerating frequency. This procedure results in different HOM spectra for each cavity. Hundreds of spectra were generated, each representing a possible cavity in reality. The six MLC cavities in CBETA have different manufacturing errors, therefore each BBU simulation in BMAD assigns each cavity one of these pre-calculated HOM spectrum. With multiple BBU simulations we therefore obtain a statistical distribution of $I_\text{th}$ of CBETA because the assigned HOM spectra will be different for each BBU simulation. 

\vbox to 1mm{}

\begin{figure}[!htb]
   \centering
   \includegraphics*[width=235pt]{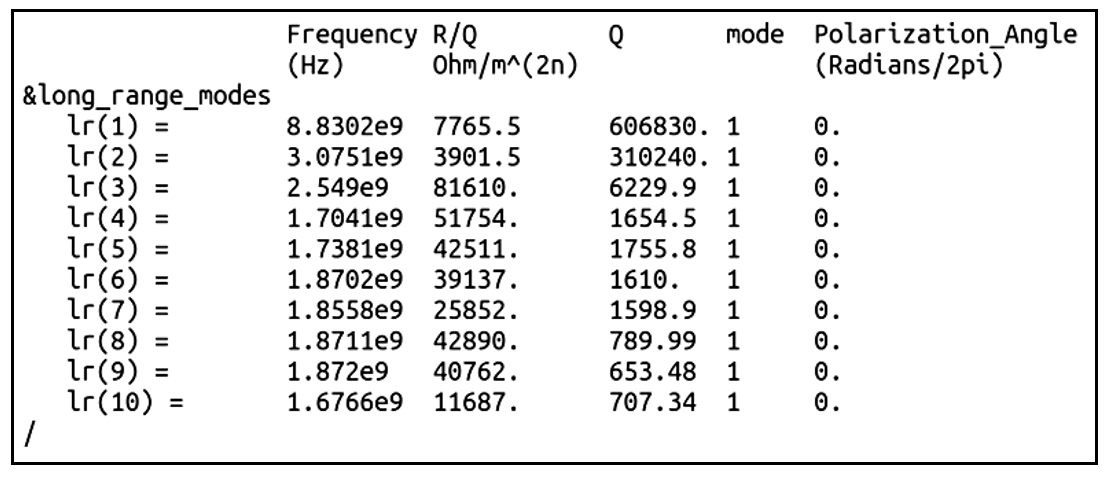}
   \caption{An example file of 10 dominant horizontal dipole-HOMs assigned to a single CBETA MLC cavity. The HOMs are simulated using CLANS2 program with $\epsilon$ = \SI{125}{\micro\meter}. Note that all the HOM frequencies are above the fundamental frequency 1.3 GHz.}
   \label{10HOMs}
\end{figure}

To save simulation time we include only the 10 most dominant transverse dipole-HOMs ($m=1$) from a pre-calculated spectrum. A dipole-HOM is considered more dominant if it has a greater figure-of-merit $\xi=(R/Q)\sqrt{Q}/f$ \cite{Valles}. 
Fig.~\ref{10HOMs} shows an example HOM assignment file with 10 dipole-HOMs for $\epsilon$ = \SI{125}{\micro\meter}. The zero polarization angles indicate that all these HOMs are horizontally polarized which give no vertical kick to the beam bunches. We include only horizontal HOMs and exclude any vertical HOMs. This is a reasonable model since the cavities have cylindrical symmetry. For the rest of this paper, HOM refers to dipole-HOM unless further specified.

\subsection{BMAD simulation detail}
The goal of BBU simulations is to find the $I_\text{th}$ for a given multipass lattice with HOMs assigned to the cavities. The BMAD program starts with a test current by injecting beam bunches into the lattice at a constant repetition rate. The initial bunches populating the lattice are given small transverse orbit offsets to allow initial excitation of the HOMs. As the bunches pass through the cavities, the momentum exchange between the bunches and the wake fields are calculated, and the HOM voltages are updated. The program records all the HOM voltages over time and periodically examine their stability. If all HOM voltages are stable over time, the test current is considered stable, and a greater current will be tested. Since the repetition rate is held constant, this is equivalent to raising the charge per bunch. In contrast, if at least one HOM voltage is unstable, the test current is regarded unstable, and a smaller current will be tested. The program typically converges to a $I_\text{th}$ within $0.1\%$ accuracy in under 30 iterations.

Since the BBU instability occurs because bunches interact with HOMs in the cavities, detailed tracking in the recirculation arc is not required. To save simulation time we usually hybridize the arc elements into an equivalent transfer matrix.
The time advantage of hybridization is one to two orders of magnitude.  

\section{BBU theory V.S BMAD Simulation}

It's important to check the validity of BMAD simulations by comparing the results to the theory predictions. A general theory of BBU has been developed in \cite{bbu_Georg_Ivan} to analytically determine the $I_\text{th}$ for a multipass lattice with multiple dipole-HOMs. Since the theory assumes thin-lens cavities, it is inaccurate to benchmark with the CBETA lattice whose cavities are each 1~m long. Instead we make a simple lattice with only thin-lens cavities and a recirculation arc with fixed optics. We will focus on four cases of which analytic formulas for $I_\text{th}$ are readily available from \cite{bbu_Georg_Ivan} and \cite{bbu_Georg_Ivan_coupled}:

\quad Case~A: One dipole-HOM with $N_p = 2$,

\quad Case~B: One dipole-HOM with $N_p = 4$,

\quad Case~C: One dipole-HOM in two different cavities 

\quad\quad\quad\quad\quad with $N_p = 2$,

\quad Case~D: Two polarized dipole-HOMs in one cavity

\quad\quad\quad\quad\quad with $N_p = 2$.

Note that $N_p$ is the number of times a bunch traverses the multipass cavity(s), and is equal to the number of recirculations plus one. For an ERL, $N_p$ must be an even number, since
each pass through a cavity for acceleration is accompanied by one for deceleration. For instance, the CBETA 1-pass lattice has $N_p=2$ and one recirculation, while the 4-pass lattice has $N_p=8$ and seven recirculations. Traditionally such an accelerator is referred to as an $N_p$/2 turn ERL. The following subsections compare the simulation results to theoretical formulas for the four cases. 

\subsection{One dipole-HOM with $N_p = 2$}

Case A is the most elementary case for BBU. Assuming that the injected current $I_0$ consists of a continuous stream of bunches with a constant charge and separated by a constant time interval $t_b$, then the time-dependent HOM voltage $V(t)$ must satisfy, for any positive integer $n$, the recursive equation \cite{bbu_Georg_Ivan}:

\begin{gather}
V(nt_b+t_r) = I_0 \frac{e}{c}t_b T_{12}\sum^\infty_{m=0}W(mt_b)V([n-m]t_b),
\end{gather}

in which $W(\tau)$ is the long range wake function characterized by the HOM parameters:

\begin{gather}
W(\tau) = \left( \frac{R}{Q} \right)_\lambda\frac{\omega_\lambda^2}{2c}e^{-(\omega_\lambda/2Q_\lambda)\tau}\sin(\omega_\lambda\tau).
\label{wake_eqn}
\end{gather}

All the related symbols are listed in Table 1, which closely follows the nomenclature used in \cite{bbu_Georg_Ivan}.

\begin{table}[H]
\centering
\begin{tabular}{|c|c|c|}
\hline
Symbol & SI Unit & Definition or Meaning  \\\hline
 $e$ & C & Elementary charge  \\\hline
 $c$ & m/s & Speed of light  \\\hline
 $t_{\text{RF}}$ & s & Fundamental RF period \\\hline
 $t_b$ & s & Injected bunch time spacing ($>t_{\text{RF}}$) \\\hline
 $t_r$ & s & Recirculation arc time (typically $>t_b$) \\\hline
 $n_r$ & - & $n_r=$ Top[$t_r/t_b$], integer \\\hline
 $\delta$ & - & $ \delta = (t_r/t_b -n_r) \in [0,1)$ \\
 & & For an ERL $ \delta \approx 0.5 $ \\\hline
 $\omega_\lambda$ & rad/s & HOM radial frequency \\\hline
 $(R/Q)_\lambda$ & $\Omega $ & normalized HOM Shunt Impedance \\\hline
 $Q_\lambda$ & -& HOM quality factor\\\hline
 $T_{12}$ & s/kg &The $T_{12}$ element of the transfer \\
 && matrix of the recirculation arc \\\hline
 $W(\tau)$ &V/mC & Long range wake function (see Eq.~(\ref{wake_eqn}))\\\hline
 $w(\delta, \omega)$ & V/mC & Sum over all wakes (see Eq.~(\ref{wake_sum}))\\\hline 
 $I_0$& A & Measured current at the injector \\\hline
 $\epsilon$ & -& $\epsilon=(\omega_\lambda/2Q_\lambda)t_b$ \\\hline
 $\kappa$ & Cs$\Omega/\text{m}^2 $& $\kappa=t_b(e/c^2)(R/Q)_\lambda(\omega_\lambda^2/2)$\\\hline
\end{tabular}
\caption{A list of important quantities in the elementary BBU theory (one dipole-HOM, $N_p=2$). $\epsilon$ is a measure of HOM decay in the time scale of $t_b$.} 
\label{caseA_fig}
\end{table}

The bunches arrive in the cavity at times $nt_b$, where they receive a transverse kick proportional to $V(nt_b)$, which then describes the transverse offset of successive bunches in the return loop. The Fourier transform 
\begin{gather}
\tilde V^{\Sigma}(\omega) = t_b \sum_{n=-\infty}^\infty V(n t_b)e^{i\omega n t_b}
\end{gather} 
is zero for every $\omega$ except when the following dispersion relation is satisfied \cite{bbu_Georg_Ivan}:  

\begin{gather}
\frac{1}{I_0} = D(\omega), \\
D(\omega) = \frac{e}{c}t_b 
T_{12} e^{i\omega n_r t_b} w(\delta, \omega).
\label{caseA_gen}
\end{gather}

The function $w(\delta, \omega)$ sums the contribution of all the long range wakes in the frequency domain: 
\begin{gather}
w(\delta, \omega) \equiv \sum_{n=0}^\infty W([n+\delta]t_b)e^{i\omega n t_b}.
\label{wake_sum}
\end{gather}

As a current, $I_0$ is a real number, and for a fixed $I_0$ there is a set of complex values of $\omega$ which satisfy Eq.~(\ref{caseA_gen}). For a small $I_0$ the voltage is stable, which means all the $\omega$ values have a negative imaginary part. If we keep increasing $I_0$, eventually instability will occur due to great excitement. This is reflected by the $\omega$s that have positive imaginary parts. At the onset of instability, one of the $\omega$ is crossing the real axis (i.e. is real), and the corresponding current $I_0$ is then the threshold current $I_\text{th}$. While it's difficult to find the $\omega$ values for a given $I_0$, it's easy to compute $D(\omega)$ given a real $\omega$. 
Most values computed will be complex and therefore correspond to an unphysical $I_0$. The largest real value of $D(\omega)$ determines the $I_\text{th}$. Due to the periodicity and symmetry in Eq.~(\ref{caseA_gen}), it is sufficient to check $\omega$ in just $[0, \pi/t_b)$ or any equivalent interval. Mathematically this can be written as:

\begin{gather}
\frac{1}{I_\text{th}} = \max_\omega[D(\omega), \text{ } D(\omega) \in \Re, \text{ }\omega \in [0,\pi/t_b)].
\label{max_dis}
\end{gather}

Eq.~(\ref{caseA_gen}), combined with Eq.~(\ref{max_dis}), is called the ``general analytic formula'' to determine the $I_\text{th}$ for case A. For a representative comparison between theory and simulation, we check how $I_\text{th}$ varies with $t_r$ while holding $t_b$ constant. The matrix element $T_{12}$ and the HOM properties are also held constant. Fig.~\ref{caseA_fig} shows the comparison result. Clearly BMAD's simulation agrees well with the general analytic formula, in both the regions with a high $I_\text{th}$ (the crest) and low $I_\text{th}$ (the trough).

If the HOM decay is insignificant on the time scale of $t_b$ ($\epsilon \ll1$), then Eq.~(\ref{caseA_gen}) can be simplified by linearization in small $\epsilon$. We call the resulting formula the ``linearized analytic formula'':
\begin{gather}
D(\omega) = -\frac{\kappa}{2}\frac{e^{i\omega t_r} T_{12}}{(\omega-\omega_\lambda)t_b+i\epsilon}.
\label{caseA_lin}
\end{gather}
Similar to the general formula, the linearized formula does not provide a closed form for $I_\text{th}$, so we still need to apply Eq.~(\ref{max_dis}) to find the $I_\text{th}$ as the smallest real $I_0$ over $\omega \in [0,\pi/t_b)$.

The usefulness of the linearized formula will be shown when $N_p >2$. On the other hand, if the HOM decay is insignificant also on the recirculation time scale ($n_r\epsilon \ll1$), then the formula can be further simplified into the ``approximate analytic formula'':
\begin{gather}
I_\text{th}=
\begin{cases}
-\frac{\epsilon}{\kappa}\frac{2}{T_{12}\sin (\omega_\lambda t_r)} \quad \quad \text{     if } T_{12}\sin (\omega_\lambda t_r) <0 \\
\frac{2}{\kappa|T_{12}|}\sqrt{\epsilon^2+(\frac{t_b}{t_r})^2 \times \text{minmod}(\omega_\lambda t_r,\pi)} \quad \text{otherwise,}
\end{cases}
\label{caseA_app}
\end{gather}
in which 
\begin{gather}
\text{minmod}(x,y) = \text{min}[\text{mod}(x,y), y-\text{mod}(x,y)].
\end{gather}

It is worth checking the applicability of the linearized and the approximate formula. This has been done in \cite{bbu_Georg_Ivan} for a case with $\epsilon = 0.00048$ and $n_r =$ 6 to 7. Their result shows great agreement with the two non-general formulas in the trough region, but not in the crest region. Here we test a new case with $\epsilon = 0.024$ and $n_r =$ 2 to 3, and the results are plotted together on Fig. 3. 

\begin{figure}[h]
	\centering
	\includegraphics[width=0.50\textwidth]{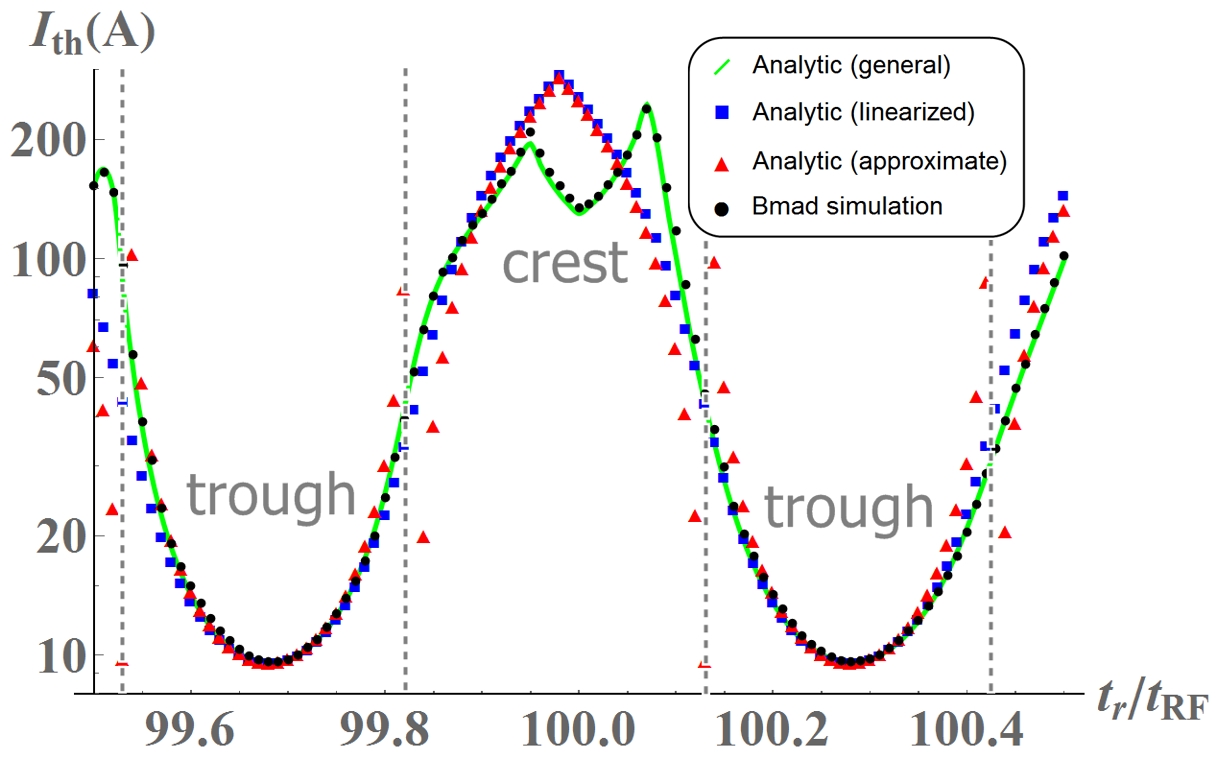}
	\caption{(Color) Comparison of the $I_\text{th}$ obtained from different analytic formulas and BMAD simulation for $N_p=2$.\\ 
	Parameters: $ct_\text{RF}=0.5$m, $t_b=50t_\text{RF}$, $\omega_\lambda/2\pi=1$GHz,  $Q=100$,  $(R/Q)_\lambda=10^4 \Omega$, $T_{12}=-10\text{m}/(1 
	\text{GeV}/c)$.}
	\label{caseA_fig}
\end{figure}

We again observe that the linearized formula agrees well with the general formula in the trough region, but the approximate formula agrees well only in a smaller region around the minimum of the trough. The inaccuracy of the approximate formula in the trough region comes from the increased value of $n_r\epsilon$.
With HOM dampers, $Q_\lambda$ is typically on the order of $10^4$, so for an ERL with continuous wave operation ($t_b = 2t_\text{RF}/N_p$, filling all the RF buckets), $\epsilon \ll 1$ is usually guaranteed. However, $n_r$ (the harmonic number of an ERL, 343 for CBETA) can be a large number depending on the recirculation lattice, so $n_r\epsilon \ll 1$ is not guaranteed. This means the approximate formula needs to be applied with caution. Note that the top case in Eq.~(\ref{caseA_app}) corresponds to the $I_\text{th}$ in the trough region, and can be rewritten as: 

\begin{gather}
I_\text{th}= \frac{-2c^2}{e(R/Q)_\lambda Q_\lambda \omega_\lambda}\frac{1}{T_{12}\sin (\omega_\lambda t_r)}.
\end{gather}

This formula has been derived in several literature regarding BBU \cite{Volkov}\cite{Yunn}\cite{Pozdeyev}. Despite its limited applicability, the formula gives us insight on how to avoid a low $I_\text{th}$. Besides suppressing the HOM quality factor $Q_\lambda$, one can also adjust the recirculation time to avoid $\sin (\omega_\lambda t_r) \approx +1$ (or $-1$) when $T_{12}$ is negative (or positive). Theoretically $I_\text{th}$ can be infinite by making $T_{12}=0$. Unfortunately this can not be achieved in general with multiple cavities and $N_p > 2$, since the $T_{12}$ between each pair of multipass cavities all needs to be zero. In reality the $T_{12}$ also depends on the length of the cavity, which will be discussed in section III-E. The strategies to improve the $I_\text{th}$ in general will be covered in section V.\\\\

\subsection{One dipole-HOM with $N_p = 4$}
In case A ($N_p=2$) we see that three analytic formulas exist: the general, linearized, and approximate formula. For a more general case with one dipole-HOM yet $N_p>2$, the general formula involves finding the maximum eigenvalue of a complex matrix \cite{bbu_Georg_Ivan}. Due to numerical difficulty we will not apply the general formula. Similar to Eq.~(\ref{caseA_lin}), the linearized formula is \cite{bbu_Georg_Ivan}:
\begin{gather}
D(\omega)= -\frac{\kappa}{2}\frac{1}{ (\omega-\omega_\lambda) t_b+i\epsilon}\sum_{J=1}^{N_p}\sum_{I=J+1}^{N_p}e^{i\omega(t_I-t_J)}T^{IJ},
\end{gather}
in which $I$ and $J$ are the cavity pass index, $(t_I-t_J)$ is the recirculation time from pass $J$ to $I$, and $T^{IJ}$ is the corresponding $T_{12}$ matrix element. To find the $I_\text{th}$ we again apply Eq.~(\ref{max_dis}), and no complex matrix is involved. The approximate formula also exists, but works only for the ``trough regions'' in which $\sum_{J=1}^{N_p}\sum_{I=J+1}^{N_p}\sin(\omega(t_I-t_J))T^{IJ} \leq 0$:

\begin{gather}
I_\text{th}= \frac{-2c^2}{e(R/Q)_\lambda Q_\lambda \omega_\lambda}\frac{1}{\sum_{J=1}^{N_p}\sum_{I=J+1}^{N_p}\sin(\omega_\lambda (t_I-t_J))T^{IJ}}.
\end{gather}

\begin{figure}[h]
	\centering
	\includegraphics[width=0.50\textwidth]{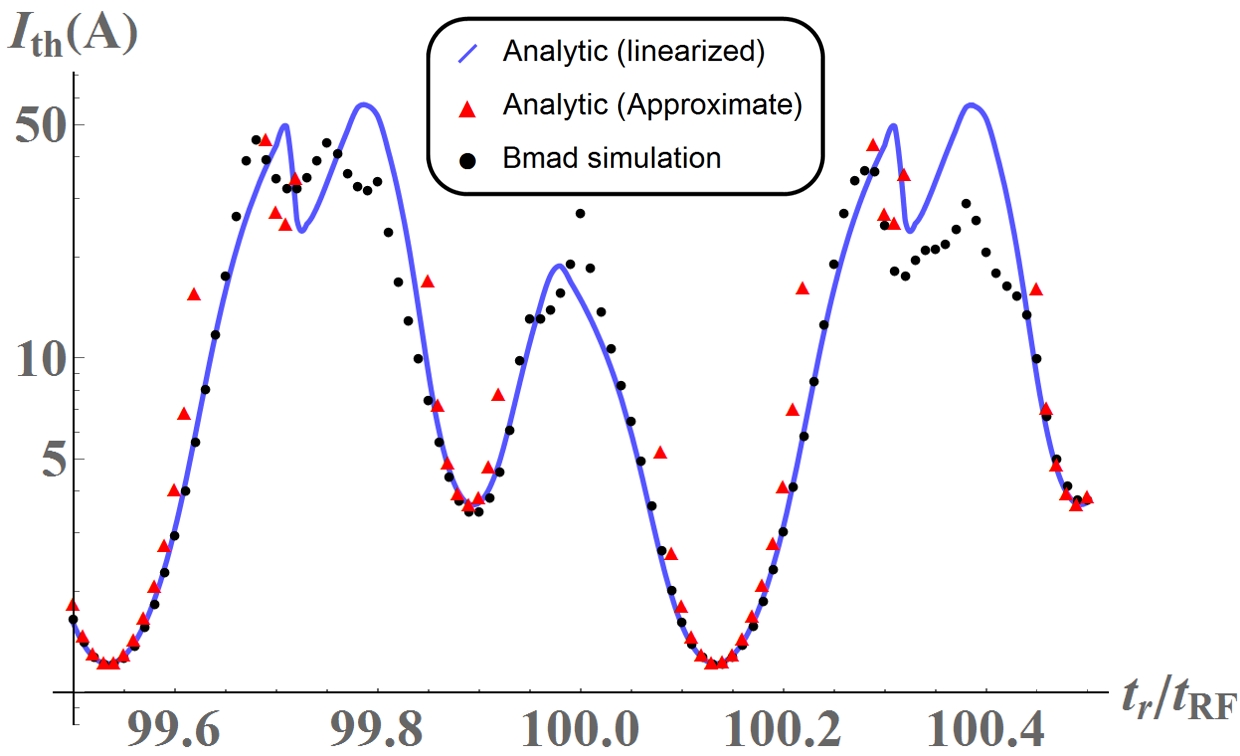}
	\caption{(Color) Comparison of the $I_\text{th}$ obtained from the linearized formula and BMAD simulation for $N_p=4$. Parameters used are the same as in Fig.~\ref{caseA_fig}, with $T_{12}=T^{IJ}$ and $t_r = t_I - t_J$ for any $I=J+1$. The trough regions are where the approximate formula (red triangles) is evaluated.}
	\label{caseB_fig}
\end{figure}

Fig.~\ref{caseB_fig} shows the comparison between BMAD simulation and the two analytic formulas. In contrast to the case with $N_p=2$ (Fig.~\ref{caseA_fig}), we now have three instead of one trough regions in one period. The number, depth, and location of the troughs depend on the signs and magnitudes of $T^{IJ}$, or the optics of multiple recirculation passes. We again observe great agreement between simulation and the linearized formula at the trough regions, and the approximate formula agrees well only around the the minimums. 

\subsection{One dipole-HOM in two different cavities with $N_p = 2$}
\begin{figure}[h]
	\centering
	\includegraphics[width=0.50\textwidth]{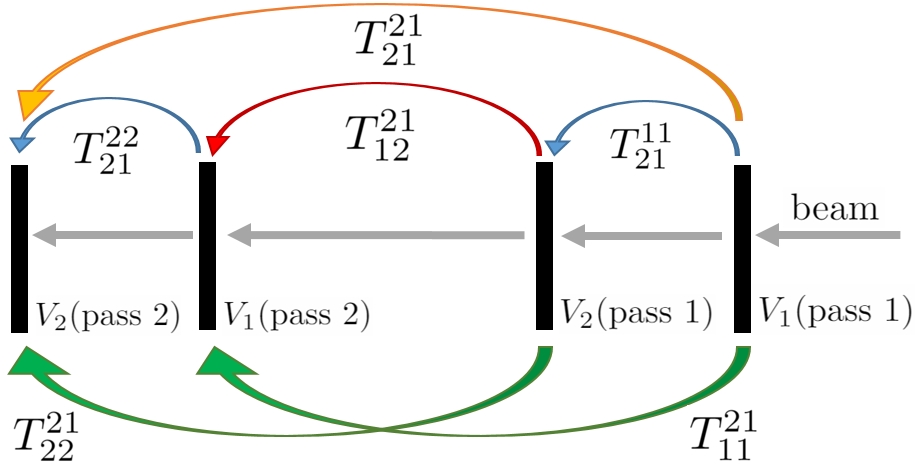}
	\caption{(Color) Illustration of the case C configuration. $V_j$ denotes the HOM of cavity $j$, and $T^{IJ}_{ij}$ is the $T_{12}$ from HOM $j$ of pass $J$ to HOM $i$ of pass $I$. Arrows with the same color indicate that the corresponding $T^{IJ}_{ij}$ are assumed the same in order to derive Eq.~(\ref{caseC_lin}).}
	\label{caseC_arrows}
\end{figure}

The complexity of this case comes from the interaction between the two HOMs via different reciculation passes. Fig.~\ref{caseC_arrows} shows all the possible ways the HOMs excite themselves and each other. For example, the HOM of cavity 1 ($V_1$) can excite itself via recirculation (via the green arrow labeled $T^{21}_{11}$). It can also excite the HOM of cavity 2 ($V_2$) in the same pass (via the blue arrows labeled $T^{11}_{21}$ for pass 1 and $T^{22}_{21}$ for pass 2).

Similar to Case B, the general formula involves calculating the eigenvalues of a complex matrix. However, the formula greatly simplifies if the two HOMs have identical characteristics, and the lattice has symmetric optics ($T^{22}_{21}=T^{11}_{21}$ and $T^{21}_{22}=T^{21}_{11}$) \cite{bbu_Georg_Ivan}:  

\begin{gather}
D(\omega)= -\frac{\kappa}{2}\frac{e^{i\omega t_r}[T^{21}_{11}\pm\sqrt{T^{21}_{12}(T^{21}_{21}+2e^{-i\omega t_r}T^{11}_{21})}]}{(\omega-\omega_\lambda)t_b+i\epsilon}.
\label{caseC_lin}
\end{gather}

\begin{figure}[h]
	\centering
	\includegraphics[width=0.5\textwidth]{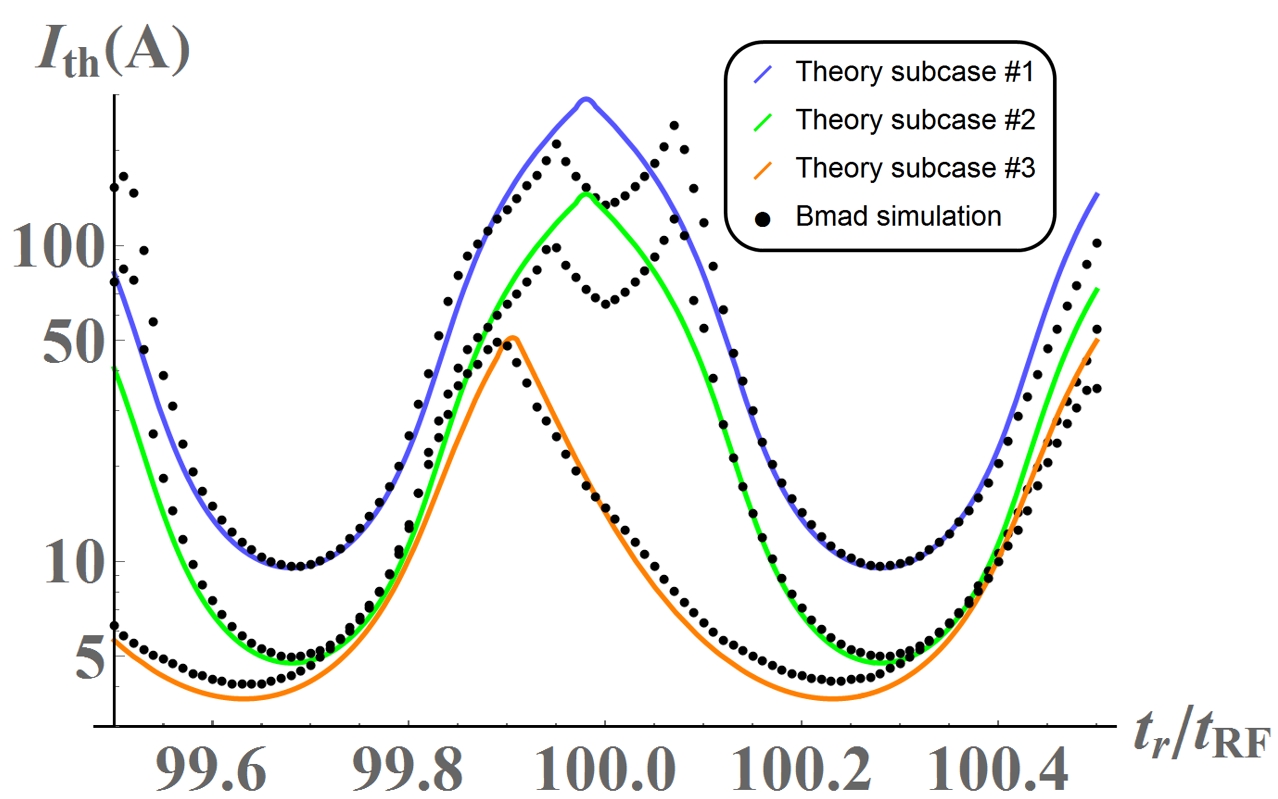}
	\caption{(Color) Comparison of the $I_\text{th}$ obtained from the linearized formula and BMAD simulation for the case C1 (top curve), C2 (middle), and C3 (bottom). The HOM properties are the same as in case A, and the optics are chosen carefully so $T^{21}_{11} = -10\text{m}/(1\text{GeV}/c)$ for all the three subcases. }
	\label{caseC_fig}
\end{figure}

Comparing to Eq.~(\ref{caseA_lin}) we see the equivalent $T_{12}$ becomes $(T^{21}_{11}\pm\sqrt{T^{21}_{12}(T^{21}_{21}+2e^{-i\omega t_r}T^{11}_{21})})$, which has two possible values for a fixed $\omega$. Since Eq.~(\ref{caseC_lin}) is a linearized formula, to find the $I_\text{th}$ we need to apply Eq.~(\ref{max_dis}) while considering both values. In general one value gives a greater $|I_0^{-1}|$, which leads to the $I_\text{th}$. Eq.~(\ref{caseC_lin}) has several peculiarities which will be explained by the following three cases with special optics, and Fig.~\ref{caseC_fig} shows the theory and simulation results for these cases. 

\begin{table}[H]
\centering
\begin{tabular}{|c|c|}
\hline
Case & Optics \\\hline
C1 &  $T^{21}_{12}=0$\\\hline
C2 &  $T^{22}_{21}=T^{11}_{21}=0$\\\hline
C3 &  $T^{21}_{21}=0$\\ \hline
\end{tabular}
\caption{The three subcases for case C with special optics.}
\end{table}

In the case C1, $T^{21}_{12}=0$, which means that the second HOM ($j=2$) can not excite the first HOM ($j=1$). This is shown clearly by the red arrow in Fig.~\ref{caseC_arrows}. Even though the first HOM can excite the second HOM in this case, there is no feedback from the second HOM. The two HOMs only feedback to themselves. The $I_\text{th}$ is therefore as large as that of with one single cavity only. Eq.~(\ref{caseC_lin}) supports this argument since the equivalent $T_{12}$ is now simply $T^{21}_{11}$, which agrees with Eq.~(\ref{caseA_lin}) in the case A. The simulation results again agree well at the trough regions, as observed for all the linearized formulas before. 

For the case C2, each HOM can still excite itself directly through $T^{21}_{ii}$ (the green arrows in Fig.~\ref{caseC_arrows}). However, the two HOMs can now excite each other via recirculation through $T^{21}_{12}$ (the red arrow) and $T^{21}_{21}$ (the orange arrow). This mutual excitation results in extra feedback, and changes the equivalent $T_{12}$ to be $(T^{21}_{11}\pm\sqrt{T^{21}_{12}T^{21}_{21}})$, which is independent of $\omega$. This means the $I_\text{th}$ occurs at the same $\omega$ as in the case C1, but the value is scaled down by a constant factor depending on $\sqrt{T^{21}_{12}T^{21}_{21}}$. The scaling effect is shown by the top two curves in Fig.~\ref{caseC_fig}. Note that if we swap the HOM index $i$ and $j$, the equivalent $T_{12}$ stays the same.

For the case C3 we have $T^{21}_{21}=0$. The two HOMs still excite other (via orange and blue arrows in Fig.~\ref{caseC_arrows}), but not symmetrically as in the case C2. The bottom curve of Fig.~\ref{caseC_fig} shows the corresponding $I_\text{th}$ profile, and the location of the trough regions clearly shifts from the two previous cases. This shift is expected due to the extra $e^{-i\omega t_r}$ term in Eq.~(\ref{caseC_lin}). The crest regions might have vanished as we choose between the two quadratic values for greater $|I_0^{-1}|$. The choice at different $t_r$ varies with on the $e^{-i\omega t_r}$ term, which allows us to stay at the trough region given by one of the two values. The overall agreement with the simulation results also supports that the crest regions, at which linearized formula typically disagrees, have vanished.  

\subsection{Two polarized dipole-HOMs in one cavity with $N_p = 2$}

All the cases discussed so far assume that the HOMs are polarized in the horizontal direction only. With cylindrical symmetry there exists a vertical HOM for each horizontal HOM, and the HOM pair has identical HOM characteristics except for the polarization angle. 
If the recirculation lattice has coupled beam optics between the two transverse phase spaces (i.e. nonzero $T_{14}$ and $T_{32}$), then the two HOMs could excite each other via recirculation. Similar to case A, we consider the simplest configuration with one cavity and $N_p=2$. For the case with $\epsilon \ll1$ and $n_r \epsilon \ll1$, the approximate formulas for the $I_\text{th}$ are \cite{bbu_Georg_Ivan_coupled}:

\begin{align}
I_\text{th}&=\min(I_{\pm}), \label{caseD_app}\\
I_{\pm}&=
\begin{cases}
  -\frac{\epsilon}{\kappa}\frac{2}{T_{\pm}\sin (\omega_\lambda t_r+\nu_{\pm})} \quad \quad \text{if it is} <0 \\
  \frac{2}{\kappa T_{\pm}}\sqrt{\epsilon^2+(\frac{t_b}{t_r})^2 \times \text{minmod}(\omega_\lambda t_r+\nu_{\pm},\pi)} \quad \text{o/w},
\end{cases} \label{caseD_app_cand}\\
T_{\pm}&e^{i\nu_{\pm}}=\frac{T_{12}+T_{34}}{2}\pm\sqrt{\left(\frac{T_{12}-T_{34}}{2}\right)^2+T_{14}T_{32}}   \label{caseD_app_T}\\
 &\text{ with } T_{\pm}, \nu_{\pm} \in \Re \text{ and } T_{\pm}>0 \nonumber.
\end{align}

Note that Eq.~(\ref{caseD_app}) is essentially Eq.~(\ref{caseA_app}) with $T_{12}$ replaced by $T_{\pm}$, and $\nu_{\pm}$ added to $\omega_\lambda t_r$. From Eq.~(\ref{caseD_app}) we see there are two candidates ($I_+$ and $I_-$) for the $I_\text{th}$, and the nature of coupling (i.e. the matrix elements in Eq.~( \ref{caseD_app_T}) determines which one is the $I_\text{th}$ at different $t_r$. We define $\Delta\nu=|\nu_+-\nu_-|$, which measures the phase shift between $I_+(t_r)$ and $I_-(t_r)$. To compare the formula with simulation results, we again focus on three cases with specified optics, listed in table III below. 

\begin{table}[H]
\centering
\begin{tabular}{|c|c|c|c|c|c|c|c|}
\hline
Case & $T_{12}$ & $T_{14}$ & $T_{32}$ & $T_{34}$& &$T_-/T_+$ &$\Delta\nu$ \\\hline
D1 &  $x$& $0$& $0$& $-x$& &1&$\pi$\\\hline
D2 &  $x$& $3x$& $-2x$& $4x$& &1&4.97\\\hline
D3 &  $x$& $(2+\sqrt{6})x$& $(-2+\sqrt{6})x$& $3x$& &13.9&$2\pi$\\ \hline
\end{tabular}
\caption{The three subcases for case D with specified optics. We set $x=-10\text{m}/(\sqrt{2}\text{GeV}/c)$, and the rest of the matrix elements are set to meet symplecticity, consistent with \cite{bbu_Georg_Ivan_coupled}. The optics for case D3 was specifically chosen to obtain $\Delta\nu = 2\pi$.}
\end{table}

\begin{figure}[h]
	\centering
	\includegraphics[width=0.5\textwidth]{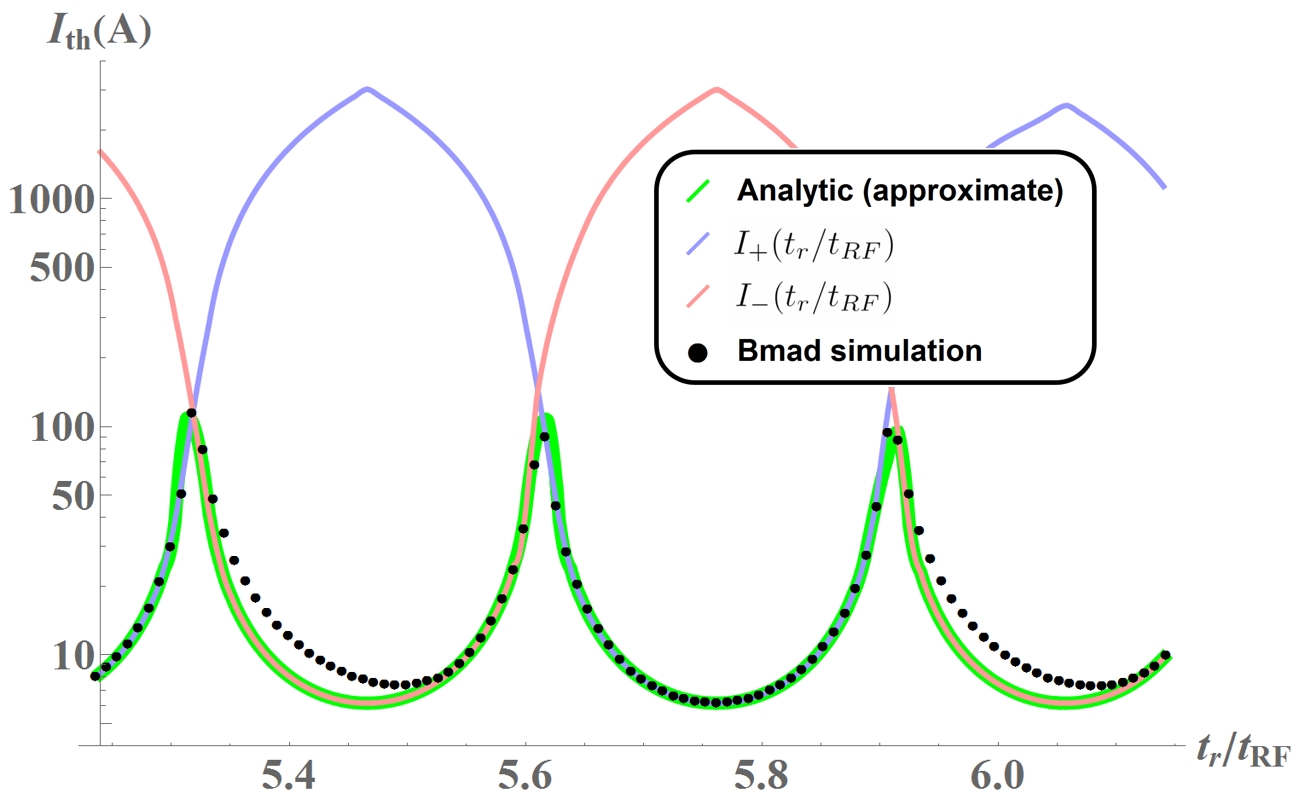}
	\caption{(Color) Comparison of the $I_\text{th}$ obtained from the approximate analytic formula (Eq.~(\ref{caseD_app})) and BMAD simulation for case D1. The two candidates for $I_\text{th}$ ($I_{\pm}$ from Eq.~(\ref{caseD_app_cand})) are also plotted.\\ 
	Parameters: $t_b=t_\text{RF}=1/1.3$ GHz, $\omega_1=\omega_2=2\pi\times2.2$ GHz,  $Q_1=Q_2=100$,  $(R/Q)_1=(R/Q)_2=10^4 \Omega$.}
	\label{caseD1_fig}
\end{figure}

Fig.~\ref{caseD1_fig} compares the $I_\text{th}$ obtained from Eq.~(\ref{caseD_app}) and BMAD simulation for the case D1. To study the behavior of coupling, the two candidates $I_{\pm}(t_r)$ are also plotted. Note that both $I_{\pm}(t_r)$ curves have distinct crest and trough regions as in case A. The two curves are $\Delta\nu=\pi$ out of phase, causing the $I_\text{th}$ to always stay at the trough regions. This is expected for two reasons. First, the lattice has no coupling ($T_{14}=T_{32}=0$), so the two HOMs do not excite each other. Mathematically we see $T_+=|T_{12}|$ and $T_-=|T_{34}|$. The second reason is about the physical difference between the trough and crest region. The trough region has lower $I_\text{th}$ because a particle with positive x offset receives positive kick in x after recirculation. In the crest region the particle instead receives a negative kick in x, resulting in a more tolerable $I_\text{th}$. Since we have $T_{12}=-T_{34}$ for subcase 1, when x motion benefits from the crest region, y motion suffers from the positive feedback at the trough region, and vice versa. The $I_\text{th}$ occurs when either x or y motion becomes unstable, not both. If we instead had $T_{12}=T_{34}$, the two candidate curves will overlap each other (in phase with equal magnitude), indicating that x and y motion are identical. In other words, without optical coupling the $I_\text{th}$ either follows Fig.~\ref{caseA_fig} (with distinct crest and trough regions) or Fig.~\ref{caseD1_fig} (with trough regions only). The BMAD simulation agrees with the approximate formula well, especially in the trough regions of $I_+(t_r)$. Reasons for the slight overestimate of $I_-(t_r)$ at the crest region are to be investigated.

\begin{figure}[h]
	\centering
	\includegraphics[width=0.5\textwidth]{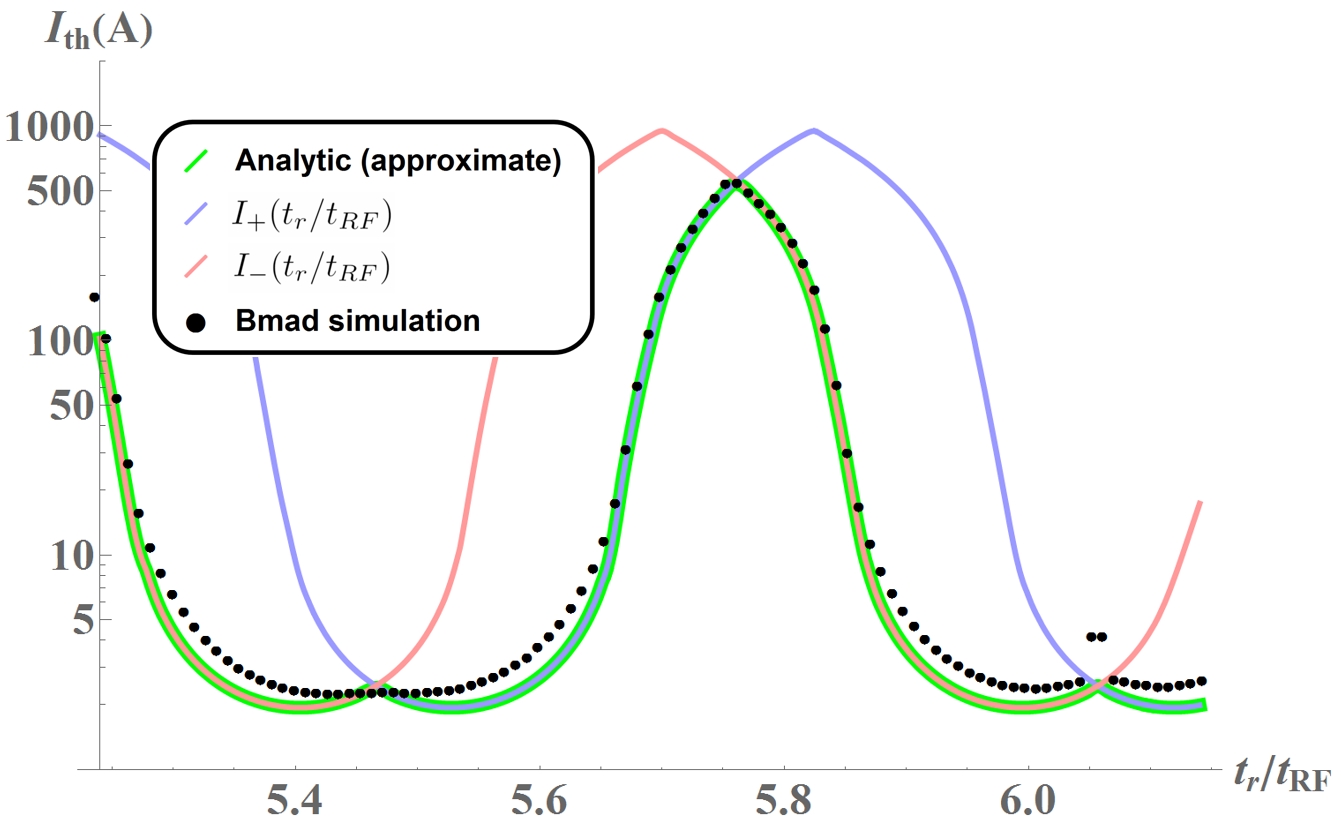}
	\caption{(Color) Comparison of the $I_\text{th}$ obtained from the approximate analytic formula (Eq.~(\ref{caseD_app})), the two candidates (Eq.~(\ref{caseD_app_cand})), and the BMAD simulation for the case D2. The parameters used are identical as in the case D1, except for the optics.}
	\label{caseD2_fig}
\end{figure}

Fig.~\ref{caseD2_fig} shows the comparison for the second subcase. The $I_\text{th}$ for this particular set of optics has been checked in \cite{bbu_Georg_Ivan_coupled} for a specific $t_r$ value, and here we check against various $t_r$ values with BMAD simulation. Similar to the case D1, case D2 has $T_+=T_-$, but the different value of $\Delta\nu$ drastically changes the $I_\text{th}$ behavior at different $t_r$. Since $\Delta\nu\neq\pi$, the crest regions of the two candidates partially overlap, giving a peak region to the $I_\text{th}$ curve. Since coupling exists now, the two transverse motions affect each other, and should not be treated independently. Around the peak, the motions together benefit from the crest regions, resulting in a greater $I_\text{th}$. Again, BMAD simulation agrees well with the approximate formula.

\begin{figure}[h]
	\centering
	\includegraphics[width=0.5\textwidth]{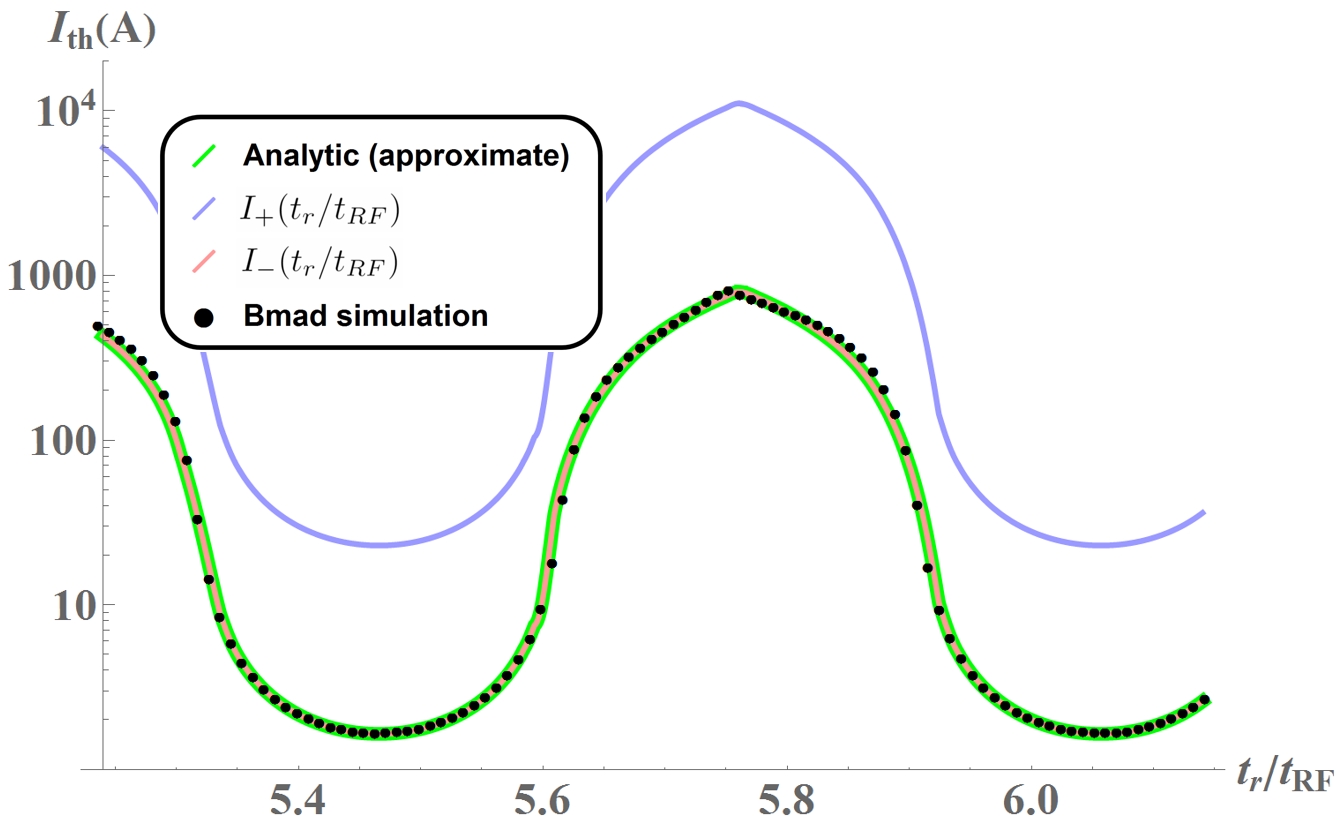}
	\caption{(Color) Comparison of the $I_\text{th}$ obtained from the approximate analytic formula (Eq.~(\ref{caseD_app})), the two candidates (Eq.~(\ref{caseD_app_cand})), and the BMAD simulation for the case D3. The parameters used are identical as in the case D1, except for the optics.}
	\label{caseD3_fig}
\end{figure}

Lastly, Fig.~\ref{caseD3_fig}  shows the comparison for the case D3. The optics are carefully chosen such that $\Delta\nu$ is $2\pi$, or equivalently zero. This causes the two candidate curves to be in phase, and the ratio $I_+(t_r)/I_-(t_r) = T_-/T_+$ remains constant. The $I_\text{th}$ curve will always follow the ``smaller'' candidate curve ($I_-(t_r)$ with our choice of optics). Recall that in the case D1 the two candidate curves would overlap (and be in phase) if $T_{12}$ and $T_{34}$ have the same sign. One might thus wonder what is the effect of coupling in the case D3. In contrast to the case D2 in which coupling changes both the magnitude and phase of the two curves, coupling here only changes their magnitude. The $I_\text{th}$ magnitude therefore entirely increases or decreases at all $t_r$ depending on the beam optics. 

The three cases above have shown that optical coupling can strongly affect the $I_\text{th}$. However, in reality it can be difficult to achieve specific optics in order to reach a high $I_\text{th}$.  For a more general case in BBU with more HOMs and $N_p >2$, neither the linearized formula nor the approximate formula exists. The general formula becomes more difficult to apply numerically, so we rely on simulation to find the $I_\text{th}$. The agreement with the analytic formulas in all the example cases makes us confident to use BMAD to calculate the $I_\text{th}$ of CBETA.

\subsection{Comment on recirculation $T_{12}$} 

\begin{figure}[h]
	\centering
	\includegraphics[width=0.5\textwidth]{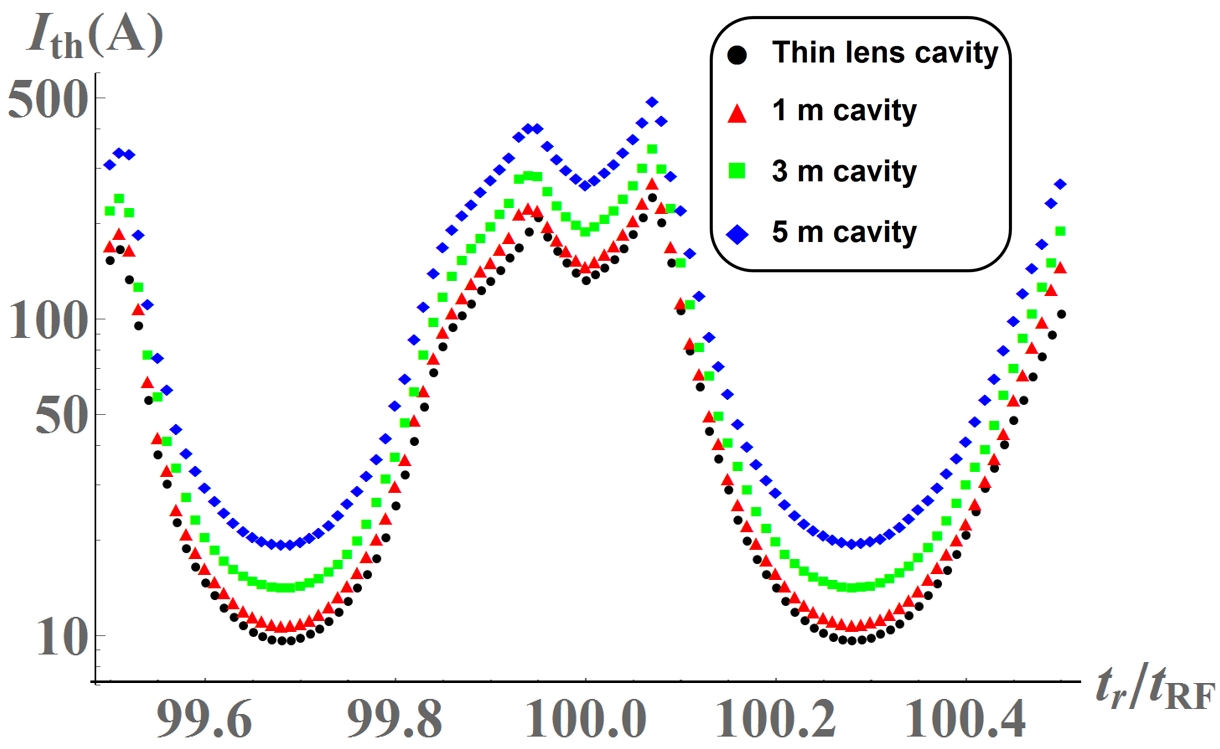}
	\caption{(Color) Comparison of the $I_\text{th}$ obtained from the BMAD simulation for case A with different cavity length. Parameters used are identical as in Fig.~\ref{caseA_fig}.}
	\label{caseE_fig}
\end{figure}

Let us refocus on the most elementary BBU case with one HOM and $N_p=2$ (Case A). Since the BBU theory derived in \cite{bbu_Georg_Ivan} assumes a thin-lens cavity, the $T_{12}$ in the formulas corresponds to the $T_{12}$ of the recirculation beamline. This is however an approximation to the reality since particles undergo transverse motion through a cavity with nonzero length. Consequently the equivalent $T_{12}$ would depend on other matrix elements ($T_{11}, T_{21}, T_{22}$, etc.) of the recirculation beamline, as well as the transfer matrix of the cavity itself. This effect is included in BMAD simulation, with the cavity transfer matrix derived in \cite{Rosen}, and the transverse HOM kick given instantly at the center of the cavity. Fig.~\ref{caseE_fig} shows the $I_\text{th}$ for case A with varying cavity length. The optics of the recirculation beam line is held constant. In our case, increasing cavity length lowers the equivalent $|T_{12}|$, resulting in a greater $I_\text{th}$ for all $t_r$. Physically this reflects the transverse focusing effect of the cavity. 

In reality the HOM kick is not instant at a specific point of the cavity, but gradual depending the time varying HOM field. A more realistic simulation would therefore integrate the field contribution from both the fundamental mode and the HOM to calculate the exact particle trajectory through the cavity. Since the HOM field depends on the interaction history of the traversed beam, the simulation can be computationally intensive. \\\\

\section{\NoCaseChange{BMAD} Simulation Result}
\label{section_CBETA_result}

As discussed, CBETA can operate in either the 1-pass or 4-pass mode, and each of the 6 MLC cavities can be assigned with a set of HOM spectrum. Hundreds of simulations with different HOM assignments were run to obtain a statistical distribution of $I_\text{th}$ for each specific CBETA design. We will investigate the following five design cases:

\quad Case (1):\quad CBETA 1-pass with $\epsilon$ = \SI{125}{\micro\meter}

\quad Case (2):\quad CBETA 4-pass with $\epsilon$ = \SI{125}{\micro\meter}

\quad Case (3):\quad CBETA 4-pass with $\epsilon$ = \SI{250}{\micro\meter}

\quad Case (4):\quad CBETA 4-pass with $\epsilon$ = \SI{500}{\micro\meter}

\quad Case (5):\quad CBETA 4-pass with $\epsilon$ = \SI{1000}{\micro\meter}

The first two cases aim to model the reality since CBETA cavities have the fabrication tolerance of $\pm$ \SI{125}{\micro\meter}. The latter three cases with greater fabrication errors are simulated for academic interest. Results of all the cases are presented as histograms in the following subsections. Note that some of the results have been presented in \cite{bbu_IPAC2018}.

\subsection{CBETA 1-Pass with $\epsilon$ = \SI{125}{\micro\meter}}

\begin{figure}[h]
	\centering
	\includegraphics[width=0.37\textwidth]{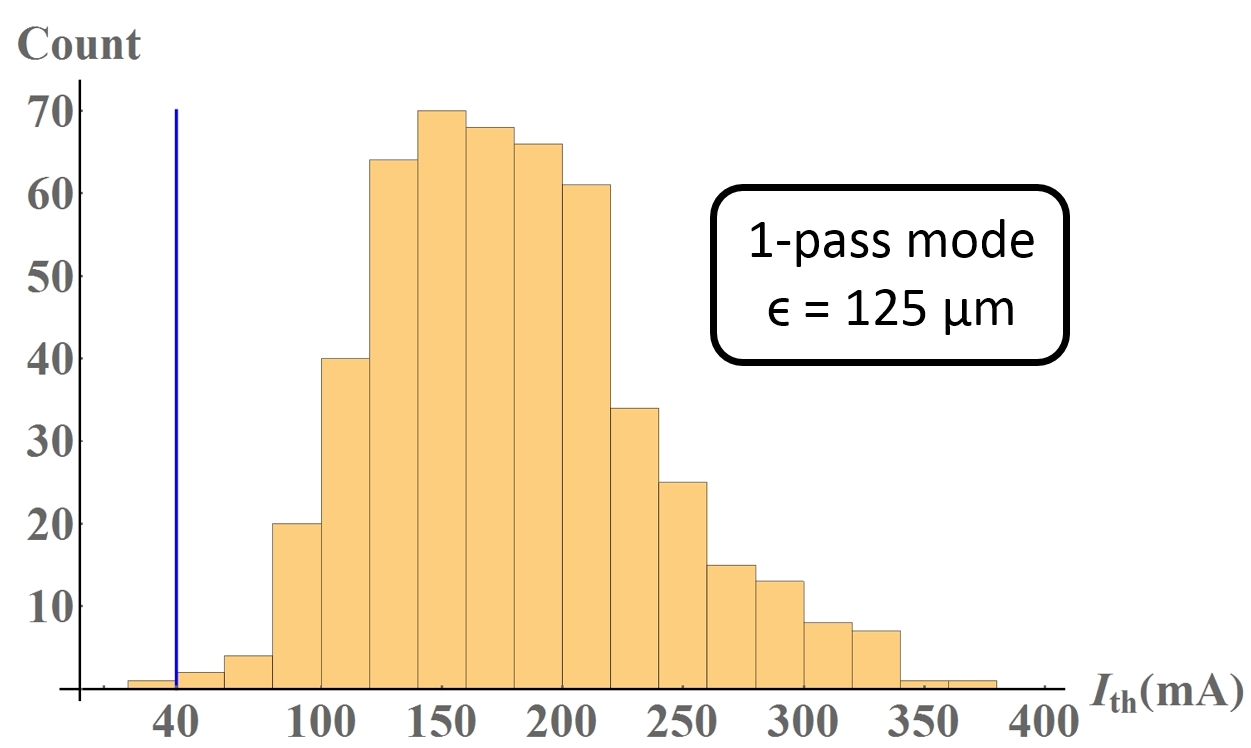}
	\caption{ 500 BBU simulation results of $I_\text{th}$ for the CBETA 1-pass lattice. Each cavity is assigned with a random set of 10 dipole HOMs ($\epsilon$ = \SI{125}{\micro\meter}). The blue line indicates the higher current goal of 40~mA.}
	\label{bbu_1pass_125um}
\end{figure}

The design current of CBETA 1-pass mode is 1~mA (the low goal) and 40~mA (the high goal). Fig.~\ref{bbu_1pass_125um} shows that all 500 simulations results exceed the lower goal of 1~mA, and only one of them is below 40~mA. This is a promising result for the CBETA 1-pass operation. We have to be unfortunate for the cavities to assume certain undesirable combinations of HOMs for the current to not reach the high goal. 

\subsection{CBETA 4-Pass with $\epsilon$ = \SI{125}{\micro\meter}}

\begin{figure}[h]
	\centering
	\includegraphics[width=0.45\textwidth]{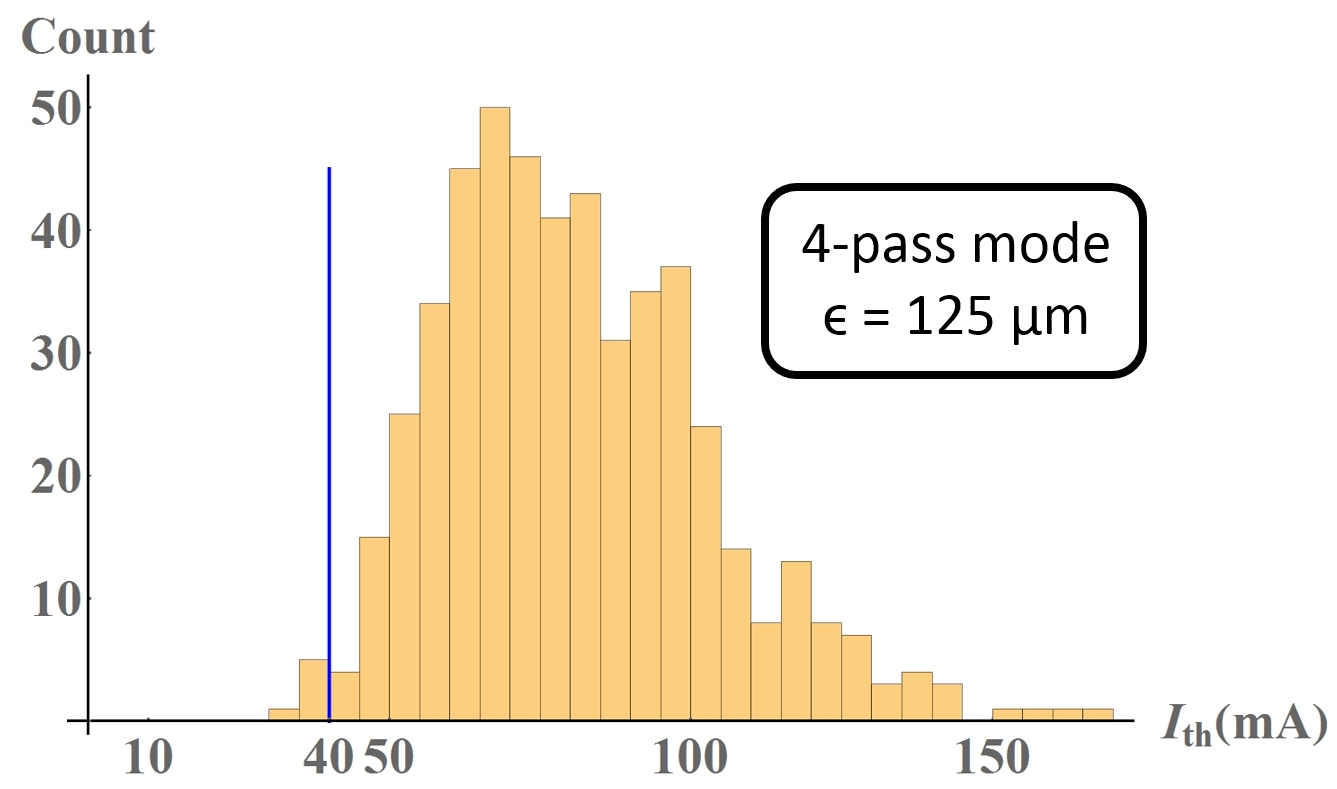}
	\caption{500 BBU simulation results of $I_\text{th}$ for the CBETA 4-pass lattice.
		Each cavity is assigned with a random set of 10 dipole HOMs ($\epsilon$ = \SI{125}{\micro\meter}). }
	\label{bbu_4pass_125um}
\end{figure}

The design current of CBETA 4-pass mode is also 1~mA and 40~mA. It's important to note that these goals refer to the injected current, so a 40~mA injected current corresponds to 80~mA for the 1-pass mode ($N_p=2$) and 320~mA for the 4-pass mode ($N_p=8$) at the MLC cavities.
Fig.~\ref{bbu_4pass_125um} shows that for the 4-pass mode, 494 out of 500 simulations exceed the 40~mA goal. This is again quite promising for the 4-pass operation, and for the few cases with undesirably low $I_\text{th}$, we will discuss the potential ways to improve them in the following section. Comparing to Fig.~\ref{bbu_1pass_125um}, the average $I_\text{th}$ for the 4-pass mode is 80.8~mA, much lower than the 179.4~mA of the 1-pass mode. This is expected from the BBU theory, since more recirculation passes allow more interaction between the HOMs and beam bunches, thus reasulting in a smaller $I_\text{th}$.

\subsection{CBETA 4-Pass with $\epsilon \geq$ \SI{250}{\micro\meter} }

\begin{figure}[h!]
	\centering
	\includegraphics[width=0.36\textwidth]{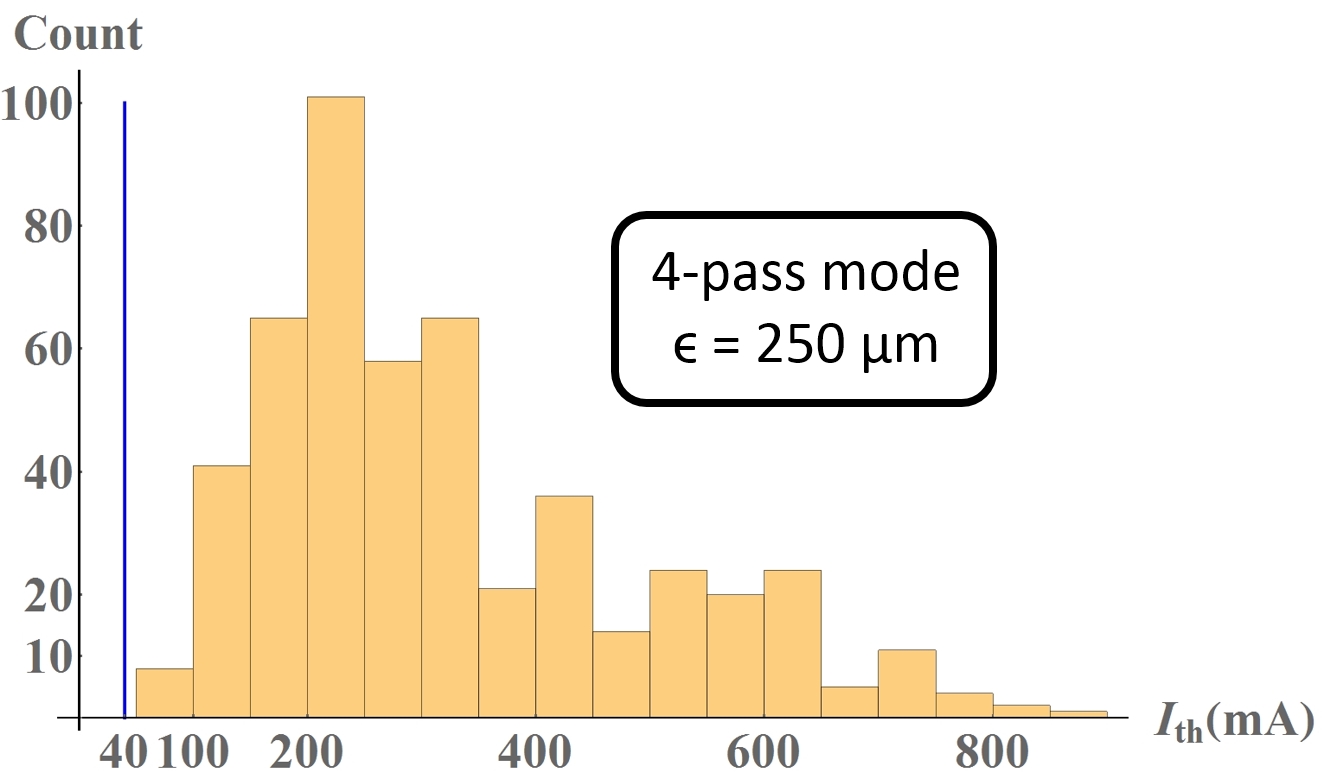}
	\caption{500 BBU simulation results of $I_\text{th}$ for the 4-pass lattice with cavity shape errors within $\epsilon$ = \SI{250}{\micro\meter}.}
	\label{bbu_4pass_250um}
\end{figure}

\begin{figure}[h!]
	\centering
	\includegraphics[width=0.36\textwidth]{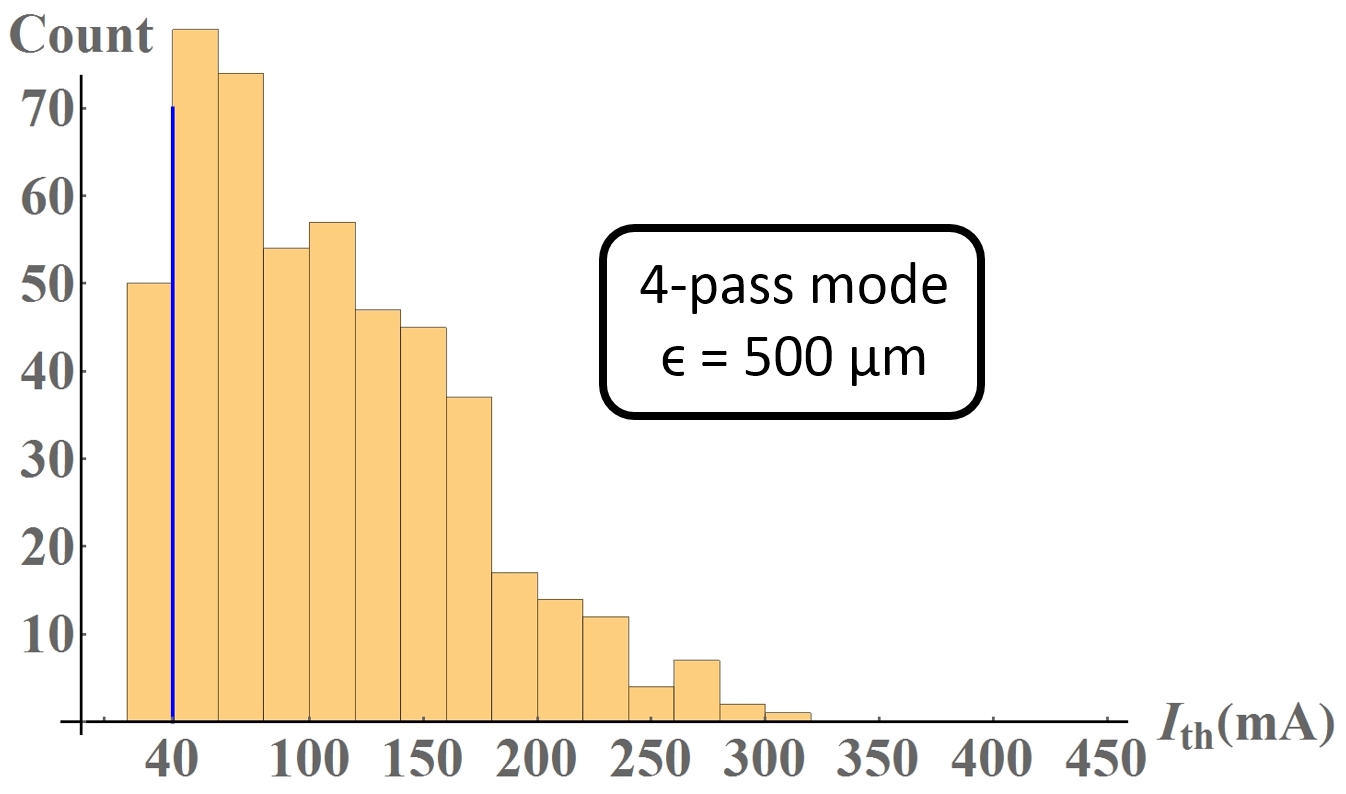}
	\caption{500 BBU simulation results of $I_\text{th}$ for the 4-pass lattice with cavity shape errors within $\epsilon$ = \SI{500}{\micro\meter}.}
	\label{bbu_4pass_500um}
\end{figure}

\begin{figure}[h!]
	\centering
	\includegraphics[width=0.36\textwidth]{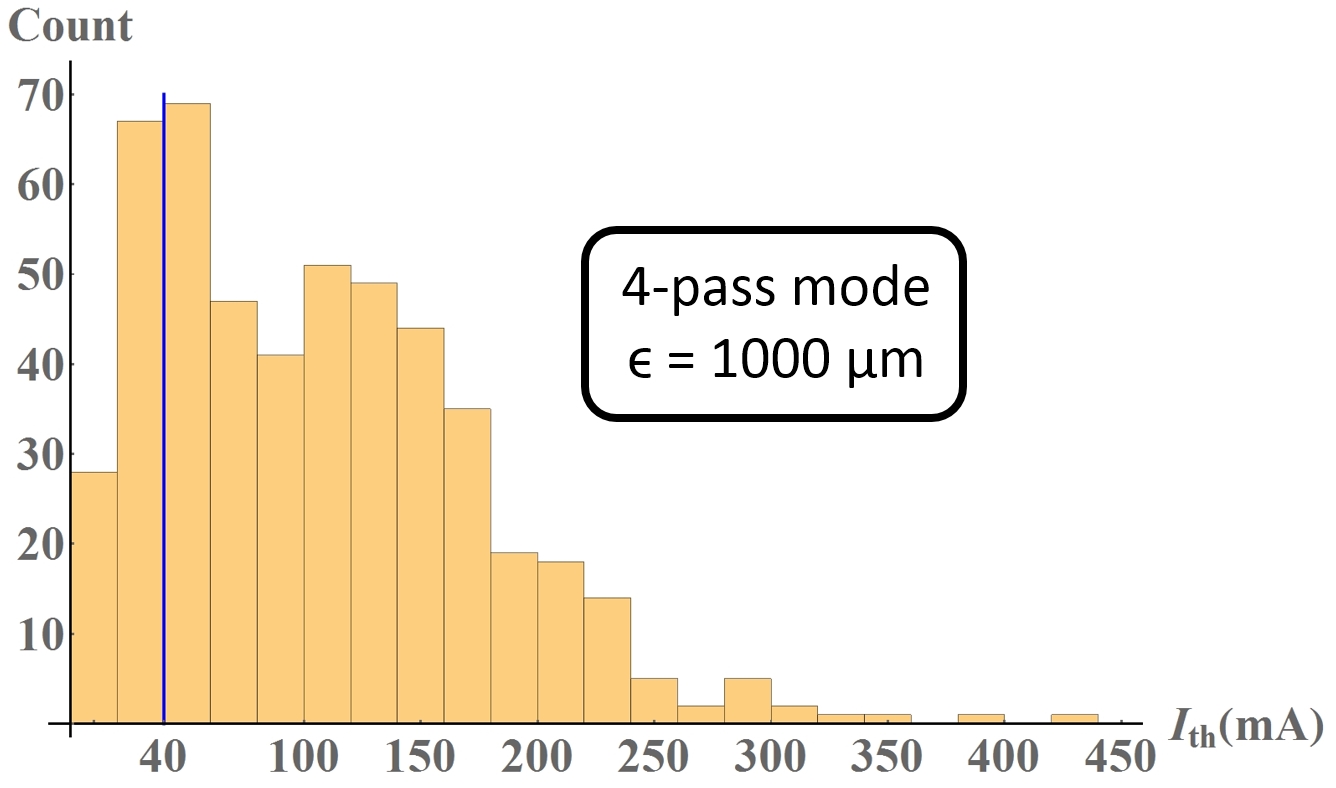}
	\caption{500 BBU simulation results of $I_\text{th}$ for the 4-pass lattice with cavity shape errors within $\epsilon$ = \SI{1000}{\micro\meter}.}
	\label{bbu_4pass_1000um}
\end{figure}

It is interesting to see how $I_\text{th}$ distribution changes with greater manufacture errors for the 4-pass lattice. Fig.~\ref{bbu_4pass_250um}, Fig.~\ref{bbu_4pass_500um}, and Fig.~\ref{bbu_4pass_1000um} show the results of 500 simulations for $\epsilon$ = \SI{250}{\micro\meter}, $\epsilon$ = \SI{500}{\micro\meter}, and $\epsilon$ = \SI{1000}{\micro\meter} respectively. For simple comparison, table 3 summarizes the statistics of all the results. For $\epsilon$ = \SI{250}{\micro\meter}, the minimum and average $I_\text{th}$ are both higher than the $\epsilon$= \SI{125}{\micro\meter} case. However, the low average $I_\text{th}$ for $\epsilon$ = \SI{1000}{\micro\meter} implies that a greater $\epsilon$ does not always improve the $I_\text{th}$.

Fundamentally greater deviation in the cavity shape results in greater spread in the HOM frequencies. This causes the HOMs across the cavities to act less coherently when kicking the beam, thus potentially increases the $I_\text{th}$. However, a greater deviation also tends to undesirably increase the $Q$ and $R/Q$ of the HOMs, which usually lowers the $I_\text{th}$. This could explain why $I_\text{th}$ statistics improves as $\epsilon$ increases from \SI{125}{\micro\meter} to \SI{250}{\micro\meter}, but deteriorates at \SI{1000}{\micro\meter}. Compensation between the frequency spread and HOM damping also implies that an optimal manufacture tolerance could exist to raise the overall $I_\text{th}$.\\\\\\

\begin{table}[h!]
\centering
\begin{tabular}{|c|c|c|c|c|c|}
\hline
CBETA mode& $\epsilon$ &$\mu(I_\text{th})$ & $\sigma(I_\text{th})$ & $\min(I_\text{th})$ & N in 500 cases\\ 
($N_p/2$)& (\SI{}{\micro\meter}) & (mA) & (mA) & (mA) & with $I_\text{th}<$ 40~mA  \\\hline
1-pass & 125 & 179.4 & 56.1 & 21.9 & 1\\\hline
4-pass & 125 & 80.8 & 22.4 & 34.4 & 6\\\hline
4-pass & 250 & 325.3 & 164.4 & 82.4 & 0\\ \hline
4-pass & 500 & 107.1 & 59.1 & 20.4 & 50\\ \hline
4-pass & 1000 & 106.6 & 69.3 & 8.8 & 95\\ \hline
\end{tabular}
\caption{Summary of the BBU $I_\text{th}$ statistics of different CBETA design cases. For the 4-pass mode, $\epsilon$ = \SI{250}{\micro\meter} generates the most satisfying $I_\text{th}$ statistics. }
\end{table}

\section{Aim for higher $I_\text{th}$}

From BBU theory we know that $I_\text{th}$ depends generally on the HOM properties ($\omega_\lambda, Q_\lambda, (R/Q)_\lambda$), the lattice properties ($t_r$ and $T_{12}$), and the injected bunch time spacing $t_b$. The previous section shows how $I_\text{th}$ can vary with different HOM spectra in the cavities. Our goal now is to study how much the $I_\text{th}$ of CBETA can improve with HOMs fixed. Based on the knowledge from BBU theory, three methods have been proposed:

\quad Method (1) Vary $t_b$

\quad Method (2) Vary phase~advance 

\quad Method (3) Introduce x-y coupling

Both the second and third method involve modifying the optics of the recirculation beamline between the pairs of multipass cavities. The idea of modifying beam optics to improve the $I_\text{th}$ was first suggested in 1980\cite{Rand}, and has been tested out at the Jefferson Lab's free electron laser \cite{Jeff}\cite{Jeff2}\cite{Jeff3}. The effect of all three methods can be simulated using BMAD, with results presented in the three following subsections.  

\subsection{Effect on $I_\text{th}$ by varying $t_b$}
\begin{figure}[!h]
	\centering
	\includegraphics[width=0.36\textwidth]{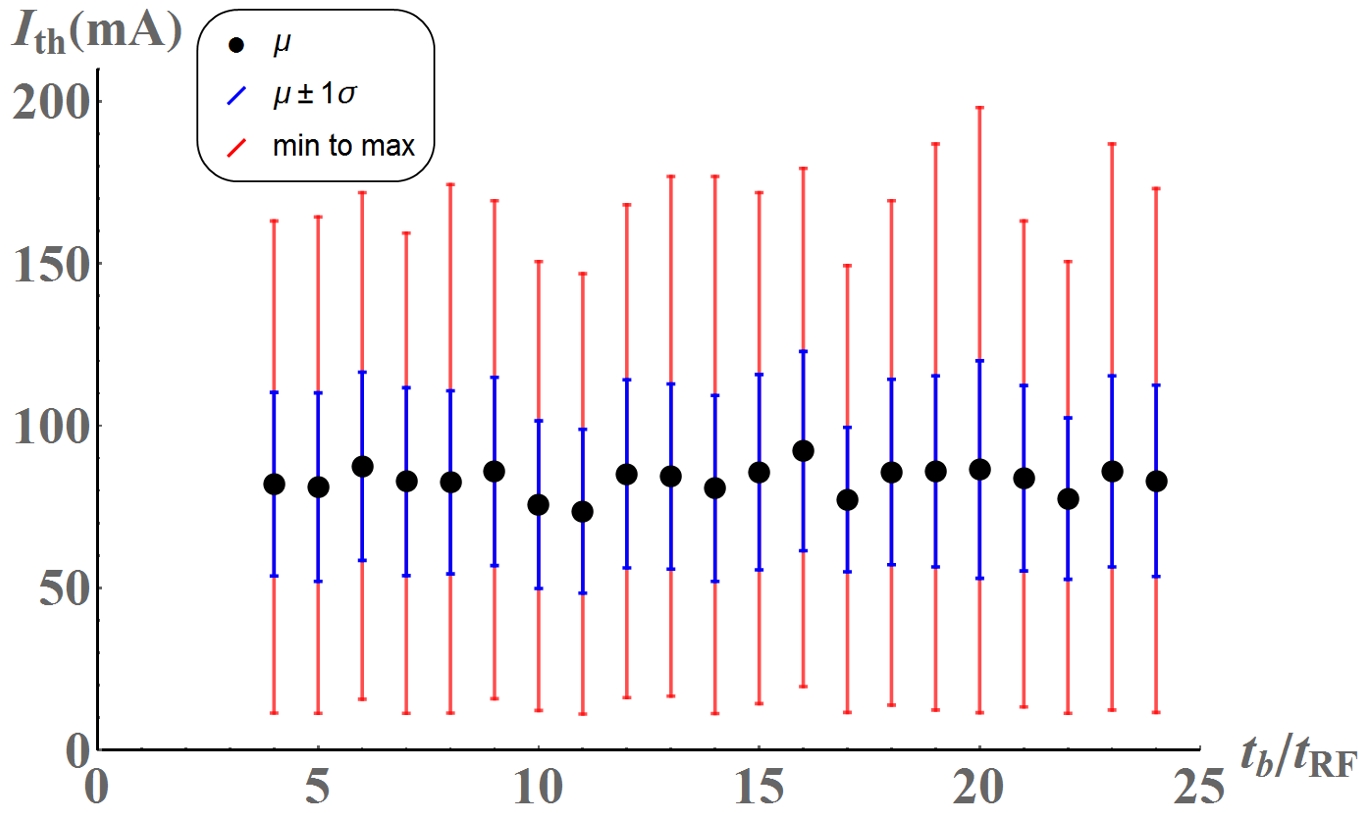}
	\caption{(Color) $I_\text{th}$ v.s $t_b/t_{\text{RF}}$ for the CBETA 4-pass lattice. For each $t_b/t_{\text{RF}}$, 500 simulations are run with different HOM assignments ($\epsilon$ = \SI{125}{\micro\meter}). The black dot marks the average $I_\text{th}$, and the blue inner line marks the $\pm 1\sigma$ range. The red outer line marks the range of the entire distribution.}
	\label{tb_fig}
\end{figure}

Eq.~(\ref{caseA_gen}) and Eq.~(\ref{wake_sum}) show that the $I_\text{th}$ depends on $t_b$ in a complicated way even for the most elementary BBU case. The dependence however vanishes in the approximate formula for the trough region (Eq.~(\ref{caseA_app})). It is interesting to investigate how $I_\text{th}$ of CBETA varies with $t_b$ using simulation. For all the bunches to see desired longitudinal acceleration, $t_b = n t_{RF}$ is required with a positive integer $n$. For all the CBETA results presented in the previous section, we have $n = N_p/2$. This corresponds to filling all the RF buckets (i.e. CW operation), and practically we would not use a smaller $n$ to avoid overlapping bunches. Fig.~\ref{tb_fig} shows the simulated $I_\text{th}$ statistics with increasing $n$ at integer steps for the 4-pass lattice ($N_p=8, \min[n]=4$). To focus on the effect of varying $t_b$ only, the 500 sets of HOM assignments are fixed. The result shows that the $I_\text{th}$ depends weakly on $t_b$, and potential improvement on $I_\text{th}$ is limited. Specifically the average $I_\text{th}$ does not change by 5$\%$. It will still be interesting to test the effect of varying $t_b$ when CBETA begins operation.

\subsection{Effect on $I_\text{th}$ by varying phase~advance}
\label{bbu_decoupled}
$I_\text{th}$ can potentially be improved by changing the phase advances (in both x and y) between the multi-pass cavities. This method equivalently changes the $T_{12}$ (and $T_{34}$) element of the transfer matrices. In the elementary case of BBU theory, smaller $T_{12}$ directly results in greater $I_\text{th}$ (Eq.~(\ref{caseA_gen})). However, with multiple cavities and HOMs, it's generally difficult to lower all the $T_{12}$ elements between different HOM pairs. To freely vary the phase advances in BMAD simulations, a zero-length lattice element is introduced right after the first pass of the MLC cavities. The element has the following 4x4 transfer matrix in the transverse phase space:

\begin{equation}
T_\text{decoupled}(\phi_{x},\phi_{y}) =
\begin{pmatrix}
   M_{x\leftarrow x} (\phi_{x}) & \boldsymbol{0}    \\
  \boldsymbol{0}  &  M_{y\leftarrow y} (\phi_{y}) 
\end{pmatrix}.
\end{equation}

Each of the 2x2 submatrix depends on the Twiss parameters ($\beta_i, \alpha_i$, and $\gamma_i$) in one transverse direction ($i = x \text{ or } y$) at the location of introduction:
\begin{equation}
M_{i \leftarrow i}(\phi) =
\begin{pmatrix}
 \cos\phi+\alpha_i\sin\phi & \beta_{i} \sin\phi \\ 
  -\gamma_{i}\sin\phi &  \cos\phi-\alpha_i\sin\phi 
\end{pmatrix}.
\end{equation}
Note that $\phi_x$ and $\phi_y$ are the additional transverse phase advances introduced by the element, and both can be chosen freely between $[0,2\pi)$. The 4x4 matrix does not introduce optical coupling between the two transverse phase spaces, and is thus named $T_\text{decoupled}$. In reality there is no physical element providing such a flexible transfer matrix, and the phase advances are changed by adjusting the quad strengths around the accelerator structure. In simulation the introduction of this matrix allows us to arbitrarily yet effectively vary the two phase advances. 

\begin{figure}[!htb]
   \centering
   \includegraphics*[width=235pt]{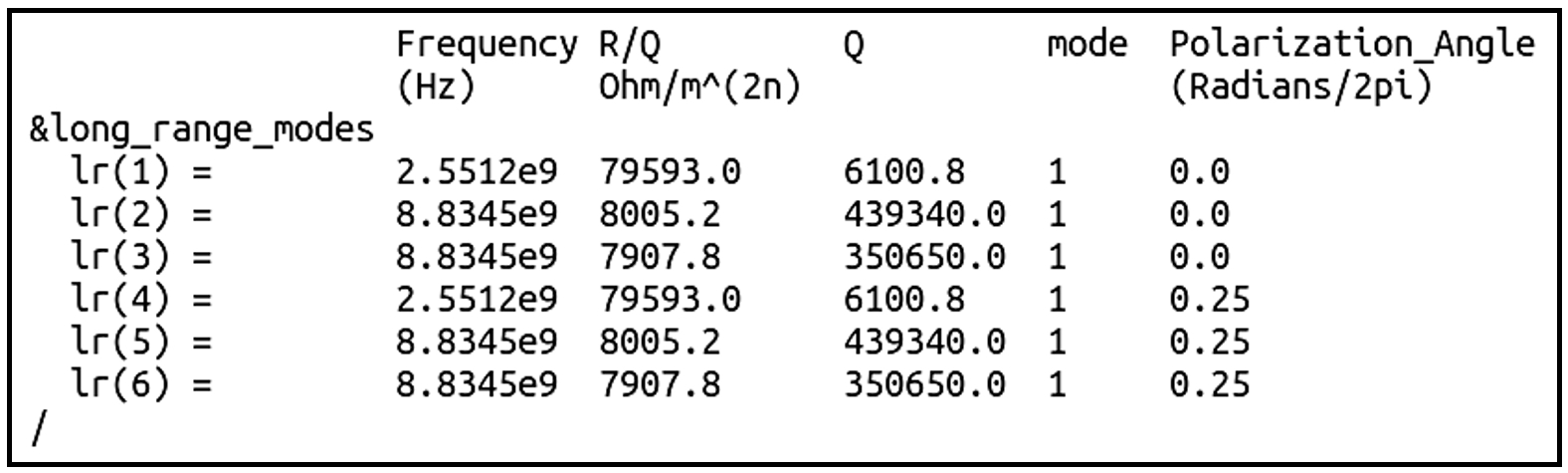}
   \caption{An example file of three dominant horizontal HOMs (the top 3) and three identical vertical HOMs (the bottom 3) assigned to a single CBETA MLC cavity. The HOMs are simulated using HTC program with $\epsilon$ = \SI{125}{\micro\meter}.}
   \label{HOM_asg_fig}
\end{figure}

To investigate how $I_\text{th}$ varies with both transverse lattice optics, we need to include vertical HOMs which give vertical kicks to the bunches. Therefore for each simulation, each cavity is assigned with three dominant $ ``\epsilon$ = \SI{125}{\micro\meter}'' horizontal HOMs and three identical vertical HOMs (polarization angle = $\pi /2$). Fig.~\ref{HOM_asg_fig} shows an example assignment to one cavity. With a fixed set of HOM assignments, the $I_\text{th}$ statistics is obtained for different choices of ($\phi_x, \phi_y$). 

\begin{figure}[h]
\centering
\includegraphics[width=0.40\textwidth,clip]{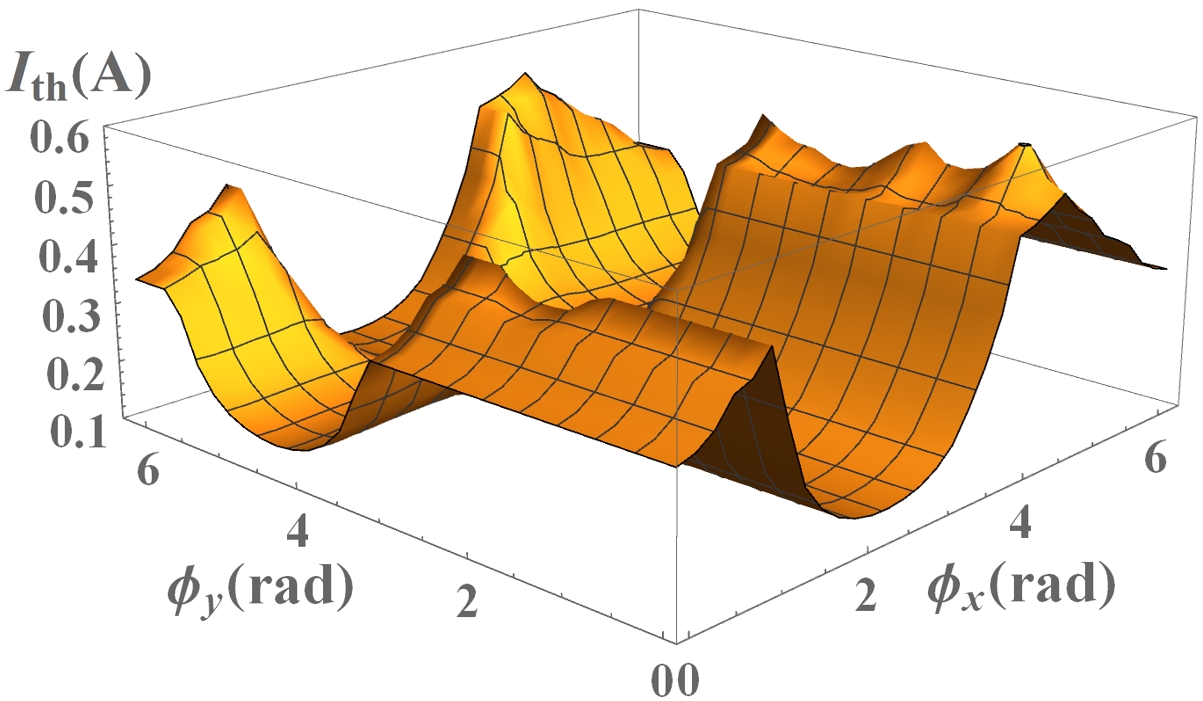}
\vskip-2mm\caption{A scan of BBU $I_\text{th}$ over the two phase advances for the CBETA 1-pass lattice. Each cavity is assigned with a random set of 3 dipole HOMs in both x and y polarization. ($\epsilon$ = \SI{125}{\micro\meter}). For this particular HOM assignment, $I_\text{th}$ ranges from 140~mA to 610~mA.}
\label{bbu_1pass_125um_decoup}
\end{figure}

\begin{figure}[h]
\centering
\includegraphics[width=0.40\textwidth,clip]{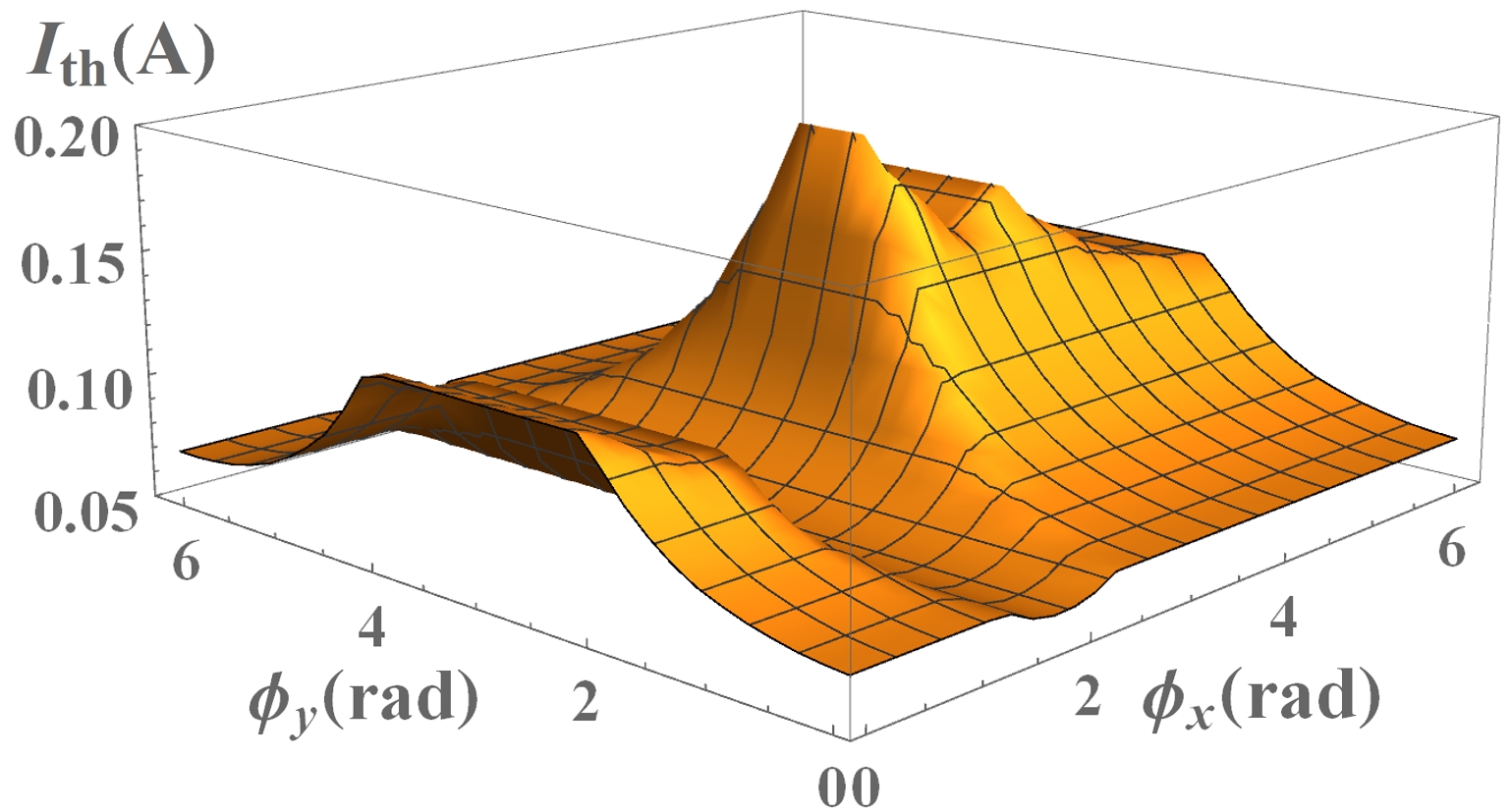}
\vskip-2mm\caption{ A scan of BBU $I_\text{th}$ over the two phase advances for the CBETA 4-pass lattice. Each cavity is assigned with a random set of 3 dipole HOMs in both x and y polarization. ($\epsilon$ = \SI{125}{\micro\meter}). For this particular HOM assignment, $I_\text{th}$ ranges from 61~mA to 193~mA.}
\label{bbu_4pass_125um_decoup}
\end{figure}

One hundred statistics were obtained for both the 1-pass and 4-pass CBETA lattice, and typical statistics are shown by  Fig.~\ref{bbu_1pass_125um_decoup} and Fig.~\ref{bbu_4pass_125um_decoup} respectively. Depending on the HOM assignment, the peak $I_\text{th}$ can reach at least 461~mA for the 1-pass mode (and 171~mA for the 4-pass mode) with an optimal choice of ($\phi_x, \phi_y$). Table IV summarizes the statistics of the peak $I_\text{th}$ with the 100 different HOM assignments. Clearly varying phase advances can be used to (significantly) improve the $I_\text{th}$. In reality the optimal set of ($\phi_x, \phi_y$) may not be achievable due to a limited number of free quadrupole magnets and strict constraints on beam optics. For CBETA however it suffices to have enough freedom to increase the $I_\text{th}$ over the design goal of 40~mA. 

Besides the promising peak $I_\text{th}$, Fig.~\ref{bbu_1pass_125um_decoup} and Fig.~\ref{bbu_4pass_125um_decoup} also show that $\phi_x$ and $\phi_y$ affect $I_\text{th}$ rather independently. That is, at certain $\phi_x$ which results in a low $I_\text{th}$ (the ``valley"), different choice of $\phi_y$ does not help increase $I_\text{th}$, and vise versa. It is also observed that $I_\text{th}$ is more sensitive to $\phi_x$, and the effect of $\phi_y$ becomes obvious mostly at the ``peak" in $\phi_x$. Physically this is expected since many lattice elements have a unit transfer matrix in the vertical phase space, and the effect of varying $T_{12}$ is more significant than $T_{34}$. In other words, HOMs with horizontal polarization are more often excited. As we will see this is no longer true when x-y coupling is introduced. 

It is also observed that the location of the valley remains almost fixed when HOM assignments are similar. Physically the valley occurs when the combination of phase-advances results in a great equivalent $T_{12}$ (or $T_{34}$) which excites the most dominant HOM. Therefore, the valley location depends heavily on which cavity has the most dominant HOM, and the simulation results agree with this observation. 

\subsection{Effect on $I_\text{th}$ with x-y coupling}
\label{bbu_coupled}
Another method potentially improves $I_\text{th}$ by introducing x-y coupling in the transverse optics, so that horizontal HOMs excite vertical motions and vise versa. This method has been shown very effective for 1-pass ERLs \cite{bbu_Georg_Ivan_coupled}. To simulate the coupling effect in BMAD simulation, a different 4x4 matrix of zero-length is again introduced right after the first pass of the LINAC:

\begin{equation}
T_\text{coupled}(\phi_{1},\phi_{2}) =
\begin{pmatrix}
  \boldsymbol{0}   & M_{x\leftarrow y} (\phi_{1})   \\
   M_{y\leftarrow x} (\phi_{2})   & \boldsymbol{0}
\end{pmatrix}.
\end{equation}

The elements of the two 2x2 submatrices $ M_{j\leftarrow i}(\phi)$ are specified using on the transverse Twiss parameters at the location of introduction:
\begin{equation}
\begin{aligned}
M_{11} &= \sqrt{\frac{\beta_{j}}{\beta_{i}}} (\cos\phi+\alpha_{i}\sin\phi) \\
M_{12} &= \sqrt{\beta_{j} \beta_{i}}\sin\phi \\
M_{21} &= \frac{1}{\sqrt{\beta_{j}\beta_{i}}}[(\alpha_{i}-\alpha_{j})\cos\phi-(1+\alpha_{i}\alpha_{j})\sin\phi] \\
M_{22} &= \sqrt{\frac{\beta_{j}}{\beta_{i}}} (\cos\phi-\alpha_{j}\sin\phi).
\end{aligned}
\end{equation}

The symplectic 4x4 matrix $T_\text{coupled}$ couples the lattice optics in the two transverse directions with two phases of free choice ($\phi_1,\phi_2$). Note the two phases are not the conventional phase advances, and can both range from 0 to 2$\pi$. 

\begin{figure}[h]
\centering
\includegraphics[width=0.40\textwidth,clip]{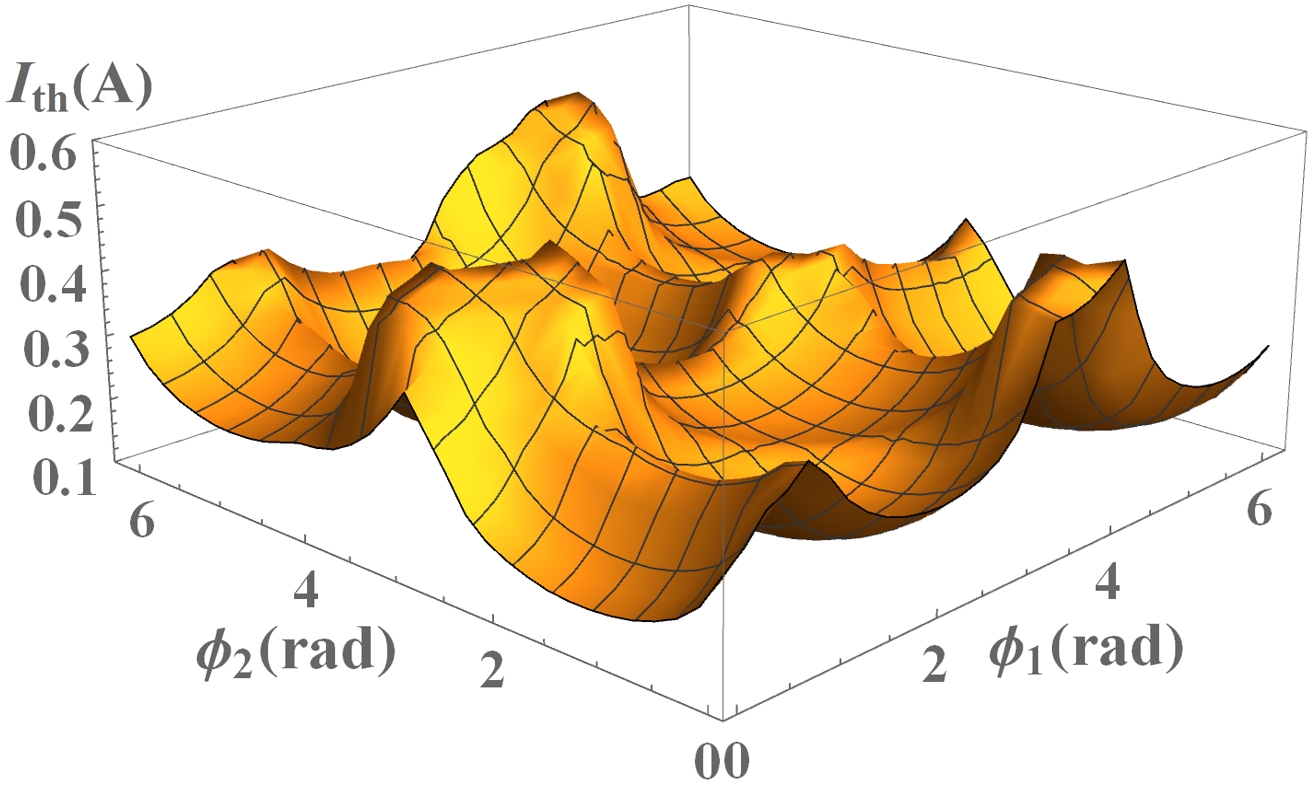}
\vskip-2mm\caption{A scan of BBU $I_\text{th}$ over the two free phases for the CBETA 1-pass lattice with x-y coupling. Each cavity is assigned with a random set of 3 dipole HOMs in both x and y polarization. ($\epsilon$ = \SI{125}{\micro\meter}). For this particular HOM assignment, $I_\text{th}$ ranges from 140~mA to 520~mA.}
\label{bbu_1pass_125um_coup}
\end{figure}

\begin{figure}[h]
\centering
\includegraphics[width=0.45\textwidth,clip]{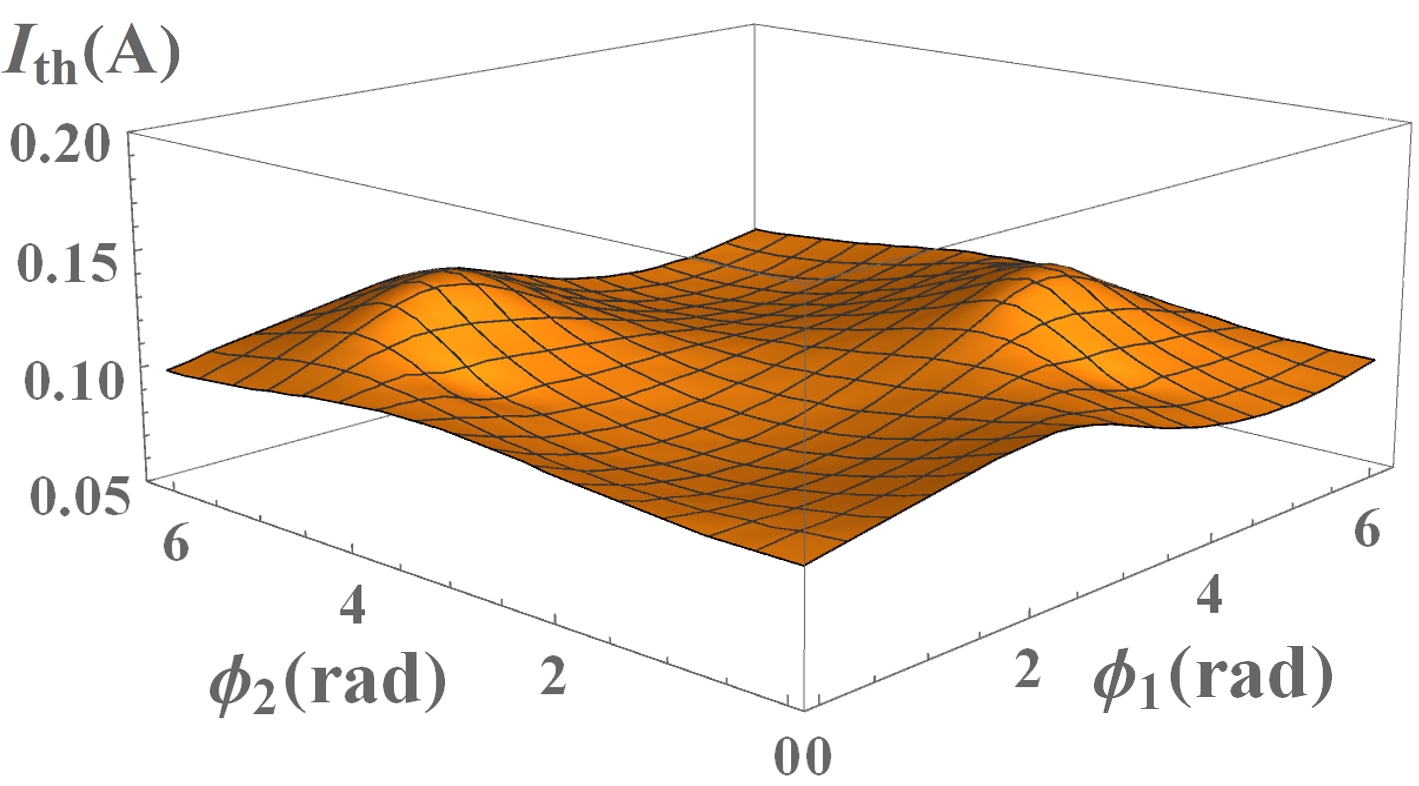}
\vskip-2mm\caption{ A scan of BBU $I_\text{th}$ over the two free phases for the CBETA 4-pass lattice with x-y coupling. Each cavity is assigned with a random set of 3 dipole HOMs in both x and y polarization. ($\epsilon$ = \SI{125}{\micro\meter}). For this particular HOM assignment, $I_\text{th}$ ranges from 89~mA to 131~mA.}
\label{bbu_4pass_125um_coup}
\end{figure}

Fig.~\ref{bbu_1pass_125um_coup} and Fig.~\ref{bbu_4pass_125um_coup} show a typical way $I_\text{th}$ varies with the two free phases for the 1-pass and 4-pass lattice respectively. Depending on the HOM assignment, the $I_\text{th}$ can reach at least 299~mA for the 1-pass mode (and 127~mA for the 4-pass mode) with an optimal choice of ($\phi_1, \phi_2$). Because the transverse optics are coupled, the two phases no longer affect $I_\text{th}$ in an independent manner. That is, there is no specific $\phi_1$ which would always result in a relatively high or low $I_\text{th}$. The two phases need to be varied together to reach the peak $I_\text{th}$. 

Similar to the case with decoupled optics, 100 statistics are run for both the 1-pass and 4-pass mode with different HOM assignments, and the statistics of the peak $I_\text{th}$ are summarized in Table IV. As expected from theory, the $I_\text{th}$ can statistically reach a higher value for the 1-pass mode than the 4-pass mode.  While introducing additional phase advances and x-y coupling both give great potential to raise the peak $I_\text{th}$ (way above the high design goal of 40~mA), the former gives more. In realty, introducing x-y coupling also requires installation of skew quadrupole magnets, and CBETA might not achieve this due to limited space. In short, varying phase advances is the most promising method to improve the $I_\text{th}$ of CBETA.

\begin{table}[h!]
\centering
\begin{tabular}{|c|c|c|c|}
\hline
&$\min(\text{peak } I_\text{th})$ & $\mu(\text{peak } I_\text{th})$ & $\max(\text{peak }I_\text{th})$ \\ 
Case (optics)&  (mA) & (mA) & (mA) \\\hline
1-pass (decoupled) & 461 & 733 & 1275 \\\hline
1-pass (coupled) & 299 & 557 & 928 \\\hline
4-pass (decoupled) & 171 & 440 & 758 \\\hline
4-pass (coupled) & 127 & 434 & 548\\ \hline
\end{tabular}
\caption{Summary of the peak $I_\text{th}$ statistics with varying transverse optics over 100 different HOM-assignments for the CBETA 1-pass and 4-pass mode. For both modes, introducing additional phase advances (decoupled optics) gives greater potential to increase $I_\text{th}$ than x-y coupling.}
\end{table}

\section{Conclusion}

In terms of the BBU threshold current ($I_\text{th}$), agreement has been found between the BBU theory and BMAD simulation for the most instructive BBU configurations. This gives us confidence in BMAD simulation for determining the $I_\text{th}$ for ERL lattices with multipass cavities and multiple HOMs, like CBETA. For the latest CBETA design lattice (both the 1-pass and 4-pass mode), simulation results show that the $I_\text{th}$ can always surpass the low design current of 1~mA, and can reach the high goal of 40~mA in over 98\% of the cases depending on the HOM spectra in the MLC cavities.\\

In reality HOM absorbers are implemented within the cavities to lower the $Q$ of the HOMs, which generally increases the $I_\text{th}$. With HOM spectra fixed, $I_\text{th}$ can still improve by adjusting the injector bunch frequency by varying the lattice optics. BMAD simulation results show that for both CBETA modes, both introducing additional phase advances and x-y coupling to the beam optics allow great improvement in the $I_\text{th}$, especially the former. Note that these results assume that the phases can be varied freely in a range 2$\pi$, while in reality the allowed values are limited by the optical constraints of the CBETA lattice. It will be interesting to test the applicability and effectiveness of these methods experimentally at CBETA.

\raggedbottom

\end{document}